\documentclass[preprint,12pt]{elsarticle}

\usepackage{graphicx}

\usepackage{lineno}

\usepackage{amsmath}
\usepackage{bm}

\usepackage{float}

\usepackage{geometry}
\usepackage{threeparttable}
\biboptions{numbers,sort&compress}

\usepackage[colorlinks, citecolor=red]{hyperref}

\journal{Elsevier}
\usepackage{CJK} 
\begin{document}

\begin{frontmatter}


\title{Grad's Distribution Function for 13 Moments based Moment Gas Kinetic Solver for Steady and Unsteady Rarefied flows: Discrete and Explicit Forms}


\author[rvt]{W.~Liu}
\author[rvtt]{Z.J.~Liu}
\author[rvt]{Z.L.~Zhang}
\author[rvt]{C.J.~Teo}
\author[rvt]{C.~Shu\corref{cor1}}
\ead{mpeshuc@nus.edu.sg}
\cortext[cor1]{Corresponding author}

\address[rvt]{Department of Mechanical Engineering, National University of Singapore, 10 Kent Ridge Crescent, Singapore 119260}
\address[rvtt]{Department of Mechanics and Aerospace Engineering, Southern University of Science and Technology, Shenzhen 518055, China}

\begin{abstract}

Efficient modeling of rarefied flow has drawn widespread interest for practical engineering applications. In the present work, we proposed the Grad's distribution function for 13 moments-based moment gas kinetic solver (G13-MGKS) and the macroscopic governing equations are derived based on the moment integral of discrete Boltzmann equation in the finite volume framework. Numerical fluxes at the cell interface related to the macroscopic variables, stress and heat flux can be reconstructed from the Boltzmann integration equation at surrounding points of the cell interface directly, so the complicated partial differential equations with tedious implementation of boundary conditions in the moment method can be avoided. Meanwhile, the explicit expression of numerical fluxes is proposed, which could release the present solver the from the discretization and numerical summation in molecular velocity space. To evaluate the Grad's distribution function for 13 moments in the present framework, the G13-MGKS with the discrete and explicit form of numerical fluxes are examined by several test cases covering the steady and unsteady rarefied flows. Numerical results indicate that the G13-MGKS could simulate continuum flows accurately and present reasonable prediction for rarefied flows at moderate Knudsen number. Moreover, the tests of computations and memory costs demonstrate that the present framework could preserve the highly efficient feature. 

\end{abstract}


\end{frontmatter}


\section{INTRODUCTION}
\label{S:1}
Rarefied effects exist in many scientific studies and industrial applications covering micro-electromechanical systems (MEMS) \cite{wang_investigation_2022,su_rarefaction_2017}, ultra-tight porous media \cite{wu_apparent_2017,liu_apparent_2020,liu_rarefaction_2018,ross-jones_conjugate_2019} and high-altitude vehicles \cite{li_gas-kinetic_2009,li_gas-kinetic_2011, chen_three-dimensional_2020}. The dimensionless parameter known as the Knudsen number is defined as the ratio of the molecular mean free path (MFP) to the spatial characteristic length to approximately estimate the magnitude of the rarefied effect. Based on the Knudsen number, the rarefied flow could be classified into four flow regimes as \cite{sone_kinetic_2002, cercignani_mathematical_1969, abdelmalik_entropy_2016}: the continuum flow regime (Kn$\leq$0.001), the slip flow regime (0.001$<$Kn$\leq$0.1), the transitional flow regime (0.1$<$Kn$\leq$10), and the free flow molecular regime (Kn$>$10). Due to the breakdown of the Continuum assumption, the challenge arises for the Navier-Stokes-Fourier (NSF) equations and conventional Computational Fluid Dynamics (CFD) modeling for simulating non-equilibrium effects \cite{bird_molecular_1994, liu_rarefaction_2018, neumann_hybrid_2014}. 

From a more fundamental standpoint, the Boltzmann equation describes the state of the gaseous system by the velocity distribution function (VDF) in all flow regimes \cite{sone_kinetic_2002, cercignani_mathematical_1969}. However, theoretical analysis of the Boltzmann equation for practical modeling remains difficult due to the intricate structure of the collision term. To address the Boltzmann equation in the numerical methods, the discretization of VDF in molecular velocity space is introduced compared to conventional CFD. Two main categories of methods could be classified, i.e., the stochastic method as well as the deterministic method \cite{zhu_unified_2019}. For the stochastic method, the set of generated particles phenomenologically simulates the collision and transport and the representative algorithm is the Direct Simulation Monte Carlo (DSMC) Method \cite{bird_molecular_1994}. Great performance has been achieved by the DSMC for solving high-speed rarefied gas flows benefiting from the adaptive properties of simulated particles in the particle space. Nevertheless, the stochastic method may suffer from the statistical noise severely for the micro-scale rarefied flow and the computations of the collision term would become time-consuming in the transient regime \cite{homolle_low-variance_2007}. 

The deterministic method, in contrast to the stochastic method, adopts discrete points to evolve the VDF in truncated molecular space and the representative algorithm is the discrete velocity method (DVM) \cite{goldstein_investigations_1989,lm_yang_improved_2019}. Within the numerical scheme of the finite volume method (FVM)\cite{jameson_numerical_nodate}, DVM adopts the upwind scheme to reconstruct the VDF on the interface. Compared with the stochastic methods, the solutions of the deterministic method are no longer affected by statistical noise and have been applied in the modeling of micro-scale rarefied flows successfully. Different from the upwind scheme in the DVM, the collision is handled simultaneously with the streaming at the cell interface in the unified gas kinetic scheme (UGKS) \cite{xu_unified_2010,xu_improved_2011, liu_unified_2015}, discrete gas kinetic scheme (DUGKS) \cite{guo_discrete_2013,guo_discrete_2015} and improved discrete velocity method (IDVM) \cite{yang_improved_2018, lm_yang_improved_2019}. So the constraints on the time step and cell size have been removed. However, the evolution and numerical integration of a sizable number of discrete points greatly increase the computations and memory consumption, causing the simulation of practical problems may become unaffordable \cite{wu_deterministic_2013}. 

Rather than evolving the time-dependent VDF, Grad proposed a truncated distribution function as the linear combinations of the moments by expanding the VDF into the Hermite polynomials of the peculiar velocity \cite{grad_kinetic_1949}. Under the increasing order of the truncation, a more accurate description of the rarefied effect could be captured. Take the third order of truncation as an example, the moments of the macroscopic variables, stress and heat flux are considered in the expression of truncated VDF, named Grad’s distribution function of 13 moments (G13) \cite{grad_statistical_1952}. The macroscopic equations of the moments can be derived with the help of truncated distribution functions to close the set of equations. The governing set of 13 moments equations, 26 moments equations and 45 moments equations have been developed. More recently, the regularized version of the moment equation has been developed from the viscoelastic regularization procedure \cite{struchtrup_regularization_2003}. Compared to DVM-based algorithms, the variables are substantially reduced in the simulation and the moment methods show good performance with much less computational cost in moderate rarefied flows \cite{gu_high-order_2009,yang_hybrid_2020,liu_apparent_2020}.

Inspired by the conservation laws and finite volume method, a novel framework of Grad’s distribution-based gas kinetic flux solver has been proposed recently \cite{liu_novel_2020}. The local solution of the Boltzmann Bhatnagar-Gross-Krook (BGK) Equation \cite{liu_numerical_2011} is utilized to reconstruct the numerical flux with the help of the Grad’s distribution function. Benefiting from the Grad’s distribution, the moment integrals in the calculation could be conducted explicitly and the G13-based gas kinetic flux solver (G13-GKFS) exhibits an efficiency close to a hundred times higher than the DVM method in previous works \cite{liu_explicit_2021}. However, the update of shear stress and heat flux are computed by moments of distribution function at cell interfaces first, and then interpolated to get the values at the cell center in the G13-GKFS. This interpolation process may lose physics especially for the unsteady states.

To combine the good feature of moment method and G13-GKFS, we proposed the G13-based moment gas kinetic solver (G13-MGKS) and the governing equations of the stress and heat flux have been derived from moment integral of the discrete Boltzmann equation in the finite volume framework. The steady-state assumption of updating the stress and heat flux in the G13-GKFS can be removed so that the present method could cover unsteady flow. Compared with complex partial differential equations with tedious implementation of boundary conditions in the moment method, the macroscopic equations related to the stress and heat flux in the present work could be updated by the evaluation of the numerical flux directly and can be conducted much easier. Besides, the reconstruction of the distribution function at the interface has been simplified to one-step calculation at surrounding points of the cell interface. The calculation of macroscopic quantities at the cell interface could be omitted compared to the G13-GKFS. Moreover, the explicit expression of numerical fluxes related to the macroscopic quantities, stress and heat flux is given in the present work, so the discretization and numerical integration in molecular velocity space can be avoided and the efficient features of the solving macroscopic equations like moment method and G13-GKFS could be preserved. 

Overall, the present framework provides a concise and efficient finite volume framework for the application of distribution functions. In the present paper, The Grad’s distribution function of 13 moments (G13) is adopted and evaluated in the present framework for a variety of rarefied flows including steady and unsteady, low velocity and supersonic flows. The paper is organized as follows: The brief introduction of kinetic theory and Boltzmann-BGK Equation are presented in Section \ref{SS:2-1}. The G13-MGKS and macroscopic equations of moment terms are proposed in Section \ref{SS:2-2}. The discrete and explicit forms of numerical fluxed are given in Section \ref{SS:2-3}. Then, the detail of the gas-surface boundary and computational Sequence are included in Sections \ref{SS:2-4} and \ref{SS:2-5}, respectively. In Section \ref{S:3}, the present method with G13 is tested by four numerical examples and the conclusion is presented in section \ref{S:4}. 

\section{METHODOLOGY}
\label{S:2}

\subsection{\emph{Kinetic theory and Boltzmann-BGK equation}}
\label{SS:2-1}

Reading from the kinetic theory, the kinetic relaxation model could be formulated in the following as \cite{chapman_mathematical_1962, liu_deformation_2021}

\begin{equation}
\frac{\partial f}{\partial t}+\bm{\xi} \cdot \nabla_{\mathbf{x}} f=\frac{g-f}{\tau},
\label{eq1}
\end{equation}

\noindent where $f(\mathbf{x}, \bm{\xi}, t)$ represent the gas velocity distribution function (VDF), which relate to the partial space $\mathbf{x}=(x, y, z)^{T}$, particle velocity space $\bm{\xi}=\left(\xi_{x}, \xi_{y}, \xi_{z}\right)^{T}$ and the time $t$. The ratio of dynamic viscosity to pressure determines the mean relaxation time, i.e., $\tau=\mu / p$. To evolve the VDF in the above Eq. (\ref{eq1}), the BGK relaxation model is considered and the equilibrium state $g(\mathbf{x}, \bm{\xi}, t)$ is given as the Maxwellian distribution

\begin{equation}
g(\mathbf{x}, \bm{\xi}, t)=\frac{\rho}{\left(2 \pi R_{g} T\right)^{D / 2}} \exp \left[-\frac{(\bm{\xi}-\mathbf{U})^{2}}{2 R_{g} T}\right],
\label{eq2}
\end{equation}

\noindent in which $\rho, \mathbf{U}=(U_{x}, U_{y}, U_{z})$ and $T$ denote the density, the macroscopic velocity and the temperature, respectively. $D$ represents the dimension value and $R_{g}$ denotes the gas constant. The macroscopic quantities $\mathbf{W}=(\rho, \rho \mathbf{U}, \rho E)^{T}$, the stress tensor $\boldsymbol{\sigma}$ and heat flux $\mathbf{q}$ can be computed associated with the moment integral of the VDF as

\begin{equation}
\mathbf{W}=(\rho, \rho \mathbf{U}, \rho E)^{T}=\langle\psi f\rangle,
\label{eq3}
\end{equation}

\begin{equation}
\boldsymbol{\sigma}=\left\langle\left(\mathbf{C C}-\delta C^{2} / 3\right) f\right\rangle,
\label{eq4}
\end{equation}

\begin{equation}
\mathbf{q}=\left\langle\mathbf{C} C^{2} f\right\rangle,
\label{eq5}
\end{equation}

\noindent where the symbol $\langle\cdot\rangle=\int_{-\infty}^{+\infty} \cdot \; d\bm{\xi}$ denotes the moment integral over the entire particle velocity space, $\psi=\left(1, \bm{\xi}, \bm{\xi}^{2}/2 \right)^{T}$ denotes the moment vector and $\mathbf{C}=\bm{\xi}-\mathbf{U}$ denotes peculiar velocities. 

To discretize Eq. (\ref{eq1}) in the framework of FVM, the cell averaged VDF $f_{i}(\mathbf{x}, \bm{\xi}, t)$ and the cell averaged macroscopic variables $\mathbf{W}_{i}(\mathbf{x}, t)$ are defined as

\begin{equation}
f_{i}(\mathbf{x}, \bm{\xi}, t)=\frac{1}{\left|\Omega_{i}\right|} \int_{\Omega_{i}} f(\mathbf{x}) \mathrm{d} \mathbf{x},
\label{eq6}
\end{equation}

\begin{equation}
\mathbf{W}_{i}(\mathbf{x}, t)=\frac{1}{\left|\Omega_{i}\right|} \int_{\Omega_{i}} \mathbf{W}(\mathbf{x}) \mathrm{d} \mathbf{x} .
\label{eq7}
\end{equation}

\noindent where the $\Omega_{i}$ represents the volume of the physical cell. With the help of the cell averaged variable, the VDF within a discretized time step $\Delta t=t^{n+1}-t^{n}$ and one discrete control cell $i$ could be obtained from the integration of Eq. (\ref{eq1}) as

\begin{equation}
f_{i}^{n+1}=f_{i}^{n}-\frac{1}{\left|\Omega_{i}\right|} \sum_{j \in N(i)}\left(\int_{t^{n}}^{t^{n+1}} \bm{\xi} \cdot \mathbf{n}_{i j} f_{i j}\left(\mathbf{x}_{i j}, \bm{\xi}, t\right) d t\right)\left|S_{i j}\right|+\int_{t^{n}}^{t^{n+1}} \frac{g_{i}-f_{i}}{\tau} d t,
\label{eq8}
\end{equation}

\noindent where $N(i)$ includes all the neighboring cells of cell $i$ and the subscript $i j$ represents the relationship between cell $i$ and the neighboring cell $j$. $\left|S_{i j}\right|$ and $\mathbf{n}_{i j}=\left({n}_{x,ij}, {n}_{y,ij}, {n}_{z,ij}\right)^{T}$ represent the area and the unit normal vector of the cell interface $i j$. By adopting the trapezoidal rule for the approximation of numerical fluxes and the collision term \cite{xu_unified_2010, grunfeld_time_2014}, the evolution of the microscopic equations becomes

\begin{equation}
\begin{aligned}
f_{i}^{n+1}=f_{i}^{n} &-\frac{\Delta t}{2} \frac{1}{\left|\Omega_{i}\right|} \sum_{j \in N(i)}\left(\bm{\xi} \cdot \mathbf{n}_{i j} f_{i j}\left(\mathbf{x}_{i j}, \bm{\xi}, t^{n+1}\right)\right)\left|S_{i j}\right|+\frac{\Delta t}{2}\left(\frac{g_{i}^{n+1}-f_{i}^{n+1}}{\tau^{n+1}}\right) \\
&-\frac{\Delta t}{2} \frac{1}{\left|\Omega_{i}\right|} \sum_{j \in N(i)}\left(\bm{\xi} \cdot \mathbf{n}_{i j} f_{i j}\left(\mathbf{x}_{i j}, \bm{\xi}, t^{n}\right)\right)\left|S_{i j}\right|+\frac{\Delta t}{2}\left(\frac{g_{i}^{n}-f_{i}^{n}}{\tau^{n}}\right),
\end{aligned}
\label{eq9}
\end{equation}

\noindent where the time step is calculated by $\Delta t=\sigma_{\mathrm{CFL}} L_{\min} / \max (|\mathbf{U}|+3 \sqrt{R_{g} T})$. $L_{\min}$ is the minimum mesh length in the discrete area and $\sigma_{\mathrm{CFL}}$ represents the Courant–Friedrichs–Lewy (CFL) number \cite{blazek_computational_2015}. Based on the conservation law in a relaxation process, the relaxation term can satisfy the compatibility condition as $\langle\psi(g-f) / \tau\rangle=0$. Further conducting the moment integral of Eq. (\ref{eq9}), the conservative form of macroscopic equations is given as:

\begin{equation}
\mathbf{W}_{i}^{n+1}=\mathbf{W}_{i}^{n}-\frac{\Delta t}{2} \frac{1}{\left|\Omega_{i}\right|} \sum_{j \in N(i)} \mathbf{n}_{i j} \cdot\left(\mathbf{F}_{i j}^{n+1}+\mathbf{F}_{i j}^{n}\right)\left|S_{i j}\right|,
\label{eq10}
\end{equation}

\noindent where the corresponding numerical fluxes are given as

\begin{equation}
\begin{aligned}
\mathbf{F}_{i j}=\left\langle\bm{\xi} \psi f_{i j}\left(\mathbf{x}_{i j}, \bm{\xi}, t\right)\right\rangle.
\end{aligned}
\label{eq11}
\end{equation}

\subsection{\emph{The G13-based moment gas kinetic solver (G13-MGKS)}}
\label{SS:2-2}

To address the numerical fluxes, the VDFs at the cell interface should be constructed first. The VDF at the cell interface $f_{ij}\left(\mathbf{x}_{i j}, \bm{\xi}, t^{n+1}\right)$ could be obtained from the Boltzmann-BGK equation along the characteristic line as

\begin{equation}
\begin{aligned}
f_{i j}\left(\mathbf{x}_{i j}, \bm{\xi}, t^{n+1}\right)-f_{0}\left(\mathbf{x}_{i j}-\bm{\xi} \Delta t, \bm{\xi}, t^{n}\right)=\frac{\Delta t}{\tau^{n}}\left(g\left(\mathbf{x}_{i j}-\bm{\xi} \Delta t, \bm{\xi}, t^{n}\right)-f_{0}\left(\mathbf{x}_{i j}-\bm{\xi} \Delta t, \bm{\xi}, t^{n}\right)\right).
\end{aligned}
\label{eq12}
\end{equation}

After a simple reformatting, Eq. (\ref{eq12}) can be rewritten as

\begin{equation}
\begin{aligned}
f_{i j}\left(\mathbf{x}_{i j}, \bm{\xi}, t^{n+1}\right)=\frac{\Delta t}{\tau^{n}} g\left(\mathbf{x}_{i j}-\bm{\xi} \Delta t, \bm{\xi}, t^{n}\right)+\left(1-\frac{\Delta t}{\tau^{n}}\right) f_{0}\left(\mathbf{x}_{i j}-\bm{\xi} \Delta t, \bm{\xi}, t^{n}\right).
\end{aligned}
\label{eq13}
\end{equation}

It is easy to notice that the VDF at the cell interface present the form of a linear combination of the equilibrium state and the initial VDF at the surrounding points of the interface $\mathbf{x}_{s}=\mathbf{x}_{i j}-\bm{\xi}\Delta t$. In the conventional DVM-type method, the initial distribution function should be obtained from the interpolation of VDF at the cell center. To release the amount of computation and memory consumed by the evolution of VDFs, Grad truncates the distribution function and expresses the unknown VDFs as an explicit function of macroscopic variables. Here, the Grad's distribution function for 13 moments (G13) \cite{grad_statistical_1952} could be given as

\begin{equation}
f^{G 13}=g\left(1+\frac{\boldsymbol{\sigma}}{2 p R_{g} T} \cdot \mathbf{C C}-\frac{\mathbf{q} \cdot \mathbf{C}}{p R_{g} T}\left(1-\frac{C^{2}}{5 R_{g} T}\right)\right).
\label{eq14}
\end{equation}

To construct the initial state of VDF $f_{i j}\left(\mathbf{x}_{i j}, \bm{\xi}, t\right)$, only the macroscopic quantities, stress tensor and heat flux $\phi=(\mathbf{W}, \boldsymbol{\sigma}, \mathbf{q})^{T}$ need be interpolated to the position $\mathbf{x}_{s}$ as

\begin{equation}
\phi\left(\mathbf{x}_{s}\right)=\left\{\begin{array}{ll}
\phi^{L}\left(\mathbf{x}_{i j}\right)-\nabla \phi\left(\mathbf{x}_{i}\right)^{n} \cdot \bm{\xi} \Delta t, & \mathbf{n}_{i j} \cdot \bm{\xi} \geq 0 \\
\phi^{R}\left(\mathbf{x}_{i j}\right)+\nabla \phi\left(\mathbf{x}_{j}\right)^{n} \cdot \bm{\xi} \Delta t, & \mathbf{n}_{i j} \cdot \bm{\xi}<0
\end{array},\right.
\label{eq15}
\end{equation}

\noindent where $\phi^{L}\left(\mathbf{x}_{i j}\right)$ and $\phi^{R}\left(\mathbf{x}_{i j}\right)$ are the values reconstructed from the cell center to both sides of the interface. The surrounding points are defined as $\mathbf{x}_{s}$. The gradient $\nabla \phi(\mathbf{x})^{n}$ is computed by the van Leer limiter directly. Since the VDF of G13 is the Maxwellian equilibrium state multiplied by a polynomial of peculiar velocity, the VDF at the cell interface $f_{ij}\left(\mathbf{x}_{i j}, \bm{\xi}, t^{n+1}\right)$ could be given by the explicit formulations as

\begin{equation}
f_{i j}\left(\mathbf{x}_{i j}, \bm{\xi}, t^{n+1}\right)=\left(1-\frac{\Delta t}{\tau^{n}}\right)g\left(\mathbf{x}_{s}\right)\left(\frac{\tau^{n}}{\tau^{n} - \Delta t}+\frac{\boldsymbol{\sigma}\left(\mathbf{x}_{s}\right)}{2 p R_{g} T} \cdot \mathbf{C}\mathbf{C}-\frac{\mathbf{q}\left(\mathbf{x}_{s}\right) \cdot \mathbf{C}}{p R_{g} T}\left(1-\frac{C^{2}}{5 R_{g} T}\right)\right).
\label{eq16}
\end{equation}

In contrast to the DVM-type method, the G13-MGKS does not need to record and evolve VDFs at cell center since the VDF could be reconstructed at the cell interface locally. Now the macroscopic variables $\mathbf{W}_{i}^{n+1}$ could be updated at the cell center by on the macroscopic equation (Eq. (\ref{eq10})). 

To update the stress tensor $\boldsymbol{\sigma}\left(\mathbf{x}_{i}\right)$ and heat flux $\mathbf{q}\left(\mathbf{x}_{i}\right)$ at the cell center, the moment integral related to the stress and heat flux could be conducted on Eq. (\ref{eq9}) as

\begin{equation}
\boldsymbol{\sigma}_{i}^{n+1}=\left(1+\frac{\Delta t}{2 \tau^{n+1}}\right)^{-1}\left[\left(1-\frac{\Delta t}{2 \tau^{n}}\right) \boldsymbol{\sigma}_{i}^{n}-\frac{\Delta t}{2} \frac{1}{\Omega_{i}} \sum_{j \in N(i)} \mathbf{n}_{i j} \cdot\left(\mathbf{G}_{i j}^{n+1}+\mathbf{G}_{i j}^{n}\right)\left|S_{i j}\right|\right],
\label{eq17}
\end{equation}

\noindent and

\begin{equation}
\mathbf{q}_{i}^{n+1}=\left(1+\frac{\Delta t}{2 \tau^{n+1}}\right)^{-1}\left[\left(1-\frac{\Delta t}{2 \tau^{n}}\right) \mathbf{q}_{i}^{n}-\frac{\Delta t}{2} \frac{1}{\Omega_{i}} \sum_{j \in N(i)} \mathbf{n}_{i j} \cdot\left(\mathbf{H}_{i j}^{n+1}+\mathbf{H}_{i j}^{n}\right)\left|S_{i j}\right|\right],
\label{eq18}
\end{equation}

\noindent where the numerical fluxes related to the stress $\mathbf{G}_{i j}$ and heat flux $\mathbf{H}_{i j}$ could be defined as

\begin{equation}
\begin{aligned}
\mathbf{G}_{i j}=\left\langle\bm{\xi}\left(\mathbf{\bar{C} \bar{C}}-\delta \bar{C}^{2} / 3\right) f_{i j}\right\rangle.
\end{aligned}
\label{eq19}
\end{equation}

\noindent and

\begin{equation}
\begin{aligned}
\mathbf{H}_{i j}=\left\langle\bm{\xi}\mathbf{\bar{C} \bar{C}}^{2} f_{i j}\right\rangle. 
\end{aligned}
\label{eq20}
\end{equation}

\noindent To associate the numerical flux related to stress and heat flux with the macroscopic variables, the peculiar velocities at the cell center $\mathbf{\bar{C}}\left(\mathbf{x}_{i}\right)=\bm{\xi}-\mathbf{U}_{i}^{n+1}$ are utilized where $\mathbf{U}_{i}^{n+1}$ is the velocity vector based on the updated macroscopic variables at the cell center $\mathbf{W}_{i}^{n+1}$.

\subsection{\emph{The discrete and explicit form of numerical fluxes in G13-MGKS}}
\label{SS:2-3}

For easy handling of the treatment at the cell interface, the quadrature of VDF $f_{k}$ in the discrete velocity space could be used to approximate the moment integral in the continuous space. By replacing the moment integration with a numerical form, the discrete VDF $f_{ij, k}$ should be reconstructed and the formulas for numerical flux change to

\begin{equation}
\begin{aligned}
&\mathbf{F}_{ij}=\left\langle \bm{\xi} \psi f_{i j}\left(\mathbf{x}_{i j}, \bm{\xi}, t\right) \right\rangle_{k},
\end{aligned}
\label{eq21}
\end{equation}

\begin{equation}
\begin{aligned}
\mathbf{G}_{i j}=\left\langle\bm{\xi}\left(\mathbf{\bar{C} \bar{C}}-\delta \bar{C}^{2} / 3\right) f_{i j}\right\rangle_{k},
\end{aligned}
\label{eq22}
\end{equation}

\noindent and

\begin{equation}
\begin{aligned}
\mathbf{H}_{i j}=\left\langle\bm{\xi}\mathbf{\bar{C} \bar{C}}^{2} f_{i j}\right\rangle_{k},
\end{aligned}
\label{eq23}
\end{equation}

\noindent where the moment integral in the numerical fluxes changes to the summation of discrete VDF with the weight function $\omega_{k}$ at velocity point $\bm{\xi}_{k}$, i.e., $\langle\psi f\rangle_{k}=\sum_{k} \omega_{k} \psi_{k} f_{k}$. Generally, Gauss-Hermite quadrature rules are preferable for low-velocity issues while Newton-Cotes quadrature rules are widely adopted in supersonic flows \cite{hu_investigation_2018}.

However, the discrete form of VDFs greatly increases the number of variables, and the memory and computations consumed by numerical integration and algebraic operations become the major \cite{liu_novel_2020}. Since the Grad's distribution function expresses the VDFs as an explicit function of macroscopic variables, the moment integral could be computed in an explicit way \cite{liu_explicit_2021}. Therefore, the numerical discretization and moment integration of VDFs in molecular velocity space can be avoided.

The explicit formulations of the numerical flux related to macroscopic variables $\mathbf{F}_{i j}$, stress $\mathbf{G}_{i j}$ and heat flux $\mathbf{H}_{i j}$ would be given in the remaining part of this section. The integral with the interval from negative infinity to zero is defined as $\langle\cdot\rangle_{<0}$ and the integral with the interval from zero to infinity is defined as $\langle\cdot\rangle_{>0}$. Take two-dimensional flow as an example, the explicit form of numerical flux related to macroscopic variables $\mathbf{F}_{i j}$ could be given as

\begin{equation}
\begin{aligned}
\mathbf{F}_{i j}=\mathbf{A}^{L}+\mathbf{A}^{R}-\Delta t\left[\nabla_{n}\left(\mathbf{A}_{n}^{L}+\mathbf{A}_{n}^{R}\right)+\nabla_{\tau}\left(\mathbf{A}_{\tau}^{L}+\mathbf{A}_{\tau}^{R}\right)\right],
\end{aligned}
\label{eq24}
\end{equation}

\noindent where the $\mathbf{A}$, $\mathbf{A}_{n}$ and $\mathbf{A}_{\tau}$ are all the integration parameters. The superscripts $L$ and $R$ represent variables at the left and right sides of the interface, respectively. The subscripts ${n}$ and ${\tau}$ represent the variables along with the normal and tangential directions, respectively. Assuming variables at the left side of the interface, the integration factor could be computed as

\begin{equation}
\begin{aligned}
\mathbf{A}^{L}(1)=\left\langle\xi_{n}^{1} \xi_{\tau}^{0} \zeta^{0} f_{i j}\right\rangle_{>0}, \mathbf{A}_{n}^{L}(1)=\left\langle\xi_{n}^{2} \xi_{\tau}^{0} \zeta^{0} f_{i j}\right\rangle_{>0}, \mathbf{A}_{\tau}^{L}(1)=\left\langle\xi_{n}^{1} \xi_{\tau}^{1} \zeta^{0} f_{i j}\right\rangle_{>0},
\end{aligned}
\label{eq25}
\end{equation}

\begin{equation}
\begin{aligned}
\mathbf{A}^{L}(2)=\mathbf{A}_{n}^{L}(1), \mathbf{A}_{n}^{L}(2)=\left\langle\xi_{n}^{3} \xi_{\tau}^{0} \zeta^{0} f_{i j}\right\rangle_{>0}, \mathbf{A}_{\tau}^{L}(2)=\left\langle\xi_{n}^{2} \xi_{\tau}^{1} \zeta^{0} f_{i j}\right\rangle_{>0},
\end{aligned}
\label{eq26}
\end{equation}

\begin{equation}
\begin{aligned}
\mathbf{A}^{L}(3)=\mathbf{A}_{\tau}^{L}(1), \mathbf{A}_{n}^{L}(3)=\mathbf{A}_{\tau}^{L}(2), \mathbf{A}_{\tau}^{L}(3)=\left\langle\xi_{n}^{1} \xi_{\tau}^{2} \zeta^{0} f_{i j}\right\rangle_{>0},
\end{aligned}
\label{eq27}
\end{equation}

\noindent and

\begin{equation}
\begin{aligned}
&\mathbf{A}^{L}(4)=\frac{1}{2}\left(\mathbf{A}_{n}^{L}(2)+\mathbf{A}_{\tau}^{L}(3)+\mathbf{A}^{L}(1)\right) \\
&\mathbf{A}_{n}^{L}(4)=\frac{1}{2}\left(\mathbf{B}_{n}^{L}(1)+\mathbf{B}_{n}^{L}(3)+\mathbf{A}^{L}(2)\right) \\
&\mathbf{A}_{\tau}^{L}(4)=\frac{1}{2}\left(\mathbf{B}_{n}^{L}(2)+\mathbf{B}_{\tau}^{L}(3)+\mathbf{A}_{\tau}^{L}(1)\right)
\end{aligned},
\label{eq28}
\end{equation}

\noindent where $\zeta$ is the phase energy to replace the $\xi_{z}$. The calculation of moment integral $\left\langle\xi_{n}^{o} \xi_{\tau}^{p} \zeta^{q} f_{i j}\right\rangle_{>0}$ could be found in \ref{A:1}. For the case of variables at the right side of the interface, the $\langle\cdot\rangle_{>0}$ could be replaced by the $\langle\cdot\rangle_{<0}$ easily.

Now the macroscopic variables $\mathbf{W}_{i}^{n+1}=(\rho_{i}^{n+1}, \rho_{i}^{n+1} {U}_{x, i}^{n+1},\rho_{i}^{n+1} {U}_{y, i}^{n+1}, \rho_{i}^{n+1} E_{i}^{n+1})^{T}$ could be updated at the cell center by the macroscopic equation (Eq. (\ref{eq10})). Then the latest velocities along the normal and tangential direction of cell interface, termed $\bar{U}_{n}$ and $\bar{U}_{\tau}$, can be calculated. 

\begin{equation}
\begin{aligned}
\begin{aligned}
&\bar{U}_{n}=U_{x, i}^{n+1} n_{x, i j}+U_{y, i}^{n+1} n_{y, i j}, \\
&\bar{U}_{\tau}=U_{y, i}^{n+1} n_{x, i j}-U_{x, i}^{n+1} n_{y, i j}.
\end{aligned}
\end{aligned}
\label{eq29}
\end{equation}

To update the independent components in the stress tensor $\sigma_{n n}$, $\sigma_{n \tau}$ and $\sigma_{\tau \tau}$, the corresponding components of $G_{n n}, G_{n \tau}$ and $G_{\tau \tau}$ in $\mathbf{G}_{i j}$ in the numerical flux could be expressed as  

\begin{equation}
\begin{aligned}
G_{n n}=\frac{1}{3}\left(2 m_{11}-m_{22}-m_{33}\right),
\end{aligned}
\label{eq30}
\end{equation}

\begin{equation}
\begin{aligned}
G_{n \tau}=m_{12},
\end{aligned}
\label{eq31}
\end{equation}

\begin{equation}
\begin{aligned}
G_{\tau \tau}=\frac{1}{3}\left(2 m_{22}-m_{11}-m_{33}\right),
\end{aligned}
\label{eq32}
\end{equation}

\noindent where the parameters related to the stress $m_{11}$, $m_{12}$, $m_{22}$ and $m_{33}$ could be given as

\begin{equation}
\begin{aligned}
m_{11} &=\mathbf{B}^{L}(1)+\mathbf{B}^{R}(1)-\Delta t\left[\nabla_{n}\left(\mathbf{B}_{n}^{L}(1)+\mathbf{B}_{n}^{R}(1)\right)+\nabla_{\tau}\left(\mathbf{B}_{\tau}^{L}(1)+\mathbf{B}_{\tau}^{R}(1)\right)\right] \\
&-2 \bar{U}_{n} \mathbf{F}_{i j}(2)+\left(\bar{U}_{n}\right)^{2}\mathbf{F}_{i j}(1),
\end{aligned},
\label{eq33}
\end{equation}

\begin{equation}
\begin{aligned}
m_{12} &=\mathbf{B}^{L}(2)+\mathbf{B}^{R}(2)-\Delta t\left[\nabla_{n}\left(\mathbf{B}_{n}^{L}(2)+\mathbf{B}_{n}^{R}(2)\right)+\nabla_{\tau}\left(\mathbf{B}_{\tau}^{L}(2)+\mathbf{B}_{\tau}^{R}(2)\right)\right] \\
&-\bar{U}_{\tau} \mathbf{F}_{i j}(2),
\end{aligned},
\label{eq34}
\end{equation}

\begin{equation}
\begin{aligned}
m_{22} &=\mathbf{B}^{L}(3)+\mathbf{B}^{R}(3)-\Delta t\left[\nabla_{n}\left(\mathbf{B}_{n}^{L}(3)+\mathbf{B}_{n}^{R}(3)\right)+\nabla_{\tau}\left(\mathbf{B}_{\tau}^{L}(3)+\mathbf{B}_{\tau}^{R}(3)\right)\right.\\
&-2 \bar{U}_{\tau} \mathbf{F}_{i j}(3)+\left(\bar{U}_{\tau}\right)^{2} \mathbf{F}_{i j}(1),
\end{aligned}
\label{eq35}
\end{equation}

\begin{equation}
\begin{aligned}
m_{33}=\mathbf{B}^{L}(4)+\mathbf{B}^{R}(4)-\Delta t\left[\nabla_{n}\left(\mathbf{B}_{n}^{L}(4)+\mathbf{B}_{n}^{R}(4)\right)+\nabla_{\tau}\left(\mathbf{B}_{\tau}^{L}(4)+\mathbf{B}_{\tau}^{R}(4)\right)\right].
\end{aligned}
\label{eq36}
\end{equation}

The formulations of introduced parameters including $\mathbf{B}$, $\mathbf{B}_{n}$ and $\mathbf{B}_{\tau}$ could be found in \ref{C:1}. 

To update the components of heat flux $q_{n}$ and $q_{\tau}$, the corresponding numerical flux related to the heat flux $\mathbf{H}_{i j}=\left(H_{n}, H_{\tau}\right)^{T}$ could be expressed as  

\begin{equation}
\begin{aligned}
H_{n}=\frac{1}{2}\left(m_{111}+m_{122}+m_{133}\right),
\end{aligned}
\label{eq37}
\end{equation}

\begin{equation}
\begin{aligned}
H_{\tau}=\frac{1}{2}\left(m_{211}+m_{222}+m_{233}\right),
\end{aligned}
\label{eq38}
\end{equation}

\noindent where the parameters related to the heat flux $m_{111}$, $m_{122}$, $m_{133}$, $m_{211}$, $m_{222}$ and $m_{233}$ could be given as

\begin{equation}
\begin{aligned}
m_{111} &=\mathbf{C}^{L}(1)+\mathbf{C}^{R}(1)-\Delta t\left[\nabla_{n}\left(\mathbf{C}_{n}^{L}(1)+\mathbf{C}_{n}^{R}(1)\right)+\nabla_{\tau}\left(\mathbf{C}_{\tau}^{L}(1)+\mathbf{C}_{\tau}^{R}(1)\right)\right] \\
&-3 \bar{U}_{n}\left(m_{11}+2 \bar{U}_{n} \mathbf{F}_{i j}(2)-\left(\bar{U}_{n}\right)^{2} \mathbf{F}_{i j}(1)\right)+3 \bar{U}_{n} \mathbf{F}_{i j}(2)-\left(\bar{U}_{n}\right)^{3} \mathbf{F}_{i j}(1),
\end{aligned}
\label{eq39}
\end{equation}

\begin{equation}
\begin{aligned}
m_{122} &=\mathbf{C}^{L}(2)+\mathbf{C}^{R}(2)-\Delta t\left[\nabla_{n}\left(\mathbf{C}_{n}^{L}(2)+\mathbf{C}_{n}^{R}(2)\right)+\nabla_{\tau}\left(\mathbf{C}_{\tau}^{L}(2)+\mathbf{C}_{\tau}^{R}(2)\right)\right] \\
&-2 \bar{U}_{\tau}\left(m_{12}+\bar{U}_{\tau} \mathbf{F}_{i j}(2)\right)-\bar{U}_{n}\left(m_{22}+2 \bar{U}_{\tau} \mathbf{F}_{i j}(3)-\left(\bar{U}_{\tau}\right)^{2} \mathbf{F}_{i j}(1)\right) \\
&+\left(\bar{U}_{\tau}\right)^{2} \mathbf{F}_{i j}(2)-2 \bar{U}_{n} \bar{U}_{\tau} \mathbf{F}_{i j}(3)+\bar{U}_{n}\left(\bar{U}_{\tau}\right)^{2} \mathbf{F}_{i j}(1),
\end{aligned}
\label{eq40}
\end{equation}

\begin{equation}
\begin{aligned}
m_{133} &=\mathbf{C}^{L}(3)+\mathbf{C}^{R}(3)-\Delta t\left[\nabla_{n}\left(\mathbf{C}_{n}^{L}(3)+\mathbf{C}_{n}^{R}(3)\right)+\nabla_{\tau}\left(\mathbf{C}_{\tau}^{L}(3)+\mathbf{C}_{\tau}^{R}(3)\right)\right] \\
&- \bar{U}_{n} m_{33},
\end{aligned}
\label{eq41}
\end{equation}

\begin{equation}
\begin{aligned}
m_{211} &=\mathbf{C}^{L}(4)+\mathbf{C}^{R}(4)-\Delta t\left[\nabla_{n}\left(\mathbf{C}_{n}^{L}(4)+\mathbf{C}_{n}^{R}(4)\right)+\nabla_{\tau}\left(\mathbf{C}_{\tau}^{L}(4)+\mathbf{C}_{\tau}^{R}(4)\right)\right] \\
&-2 \bar{U}_{n}\left(m_{12}+\bar{U}_{\tau} \mathbf{F}_{i j}(2)\right)-\bar{U}_{\tau}\left(m_{11}+2 \bar{U}_{n} \mathbf{F}_{i j}(2)-\left(\bar{U}_{n}\right)^{2} \mathbf{F}_{i j}(1)\right) \\
&+\left(\bar{U}_{n}\right)^{2} \mathbf{F}_{i j}(3)+2 \bar{U}_{n} \bar{U}_{\tau} \mathbf{F}_{i j}(2)-\left(\bar{U}_{n}\right)^{2} \bar{U}_{\tau} \mathbf{F}_{i j}(1),
\end{aligned}
\label{eq42}
\end{equation}

\begin{equation}
\begin{aligned}
m_{222} &=\mathbf{C}^{L}(5)+\mathbf{C}^{R}(5)-\Delta t\left[\nabla_{n}\left(\mathbf{C}_{n}^{L}(5)+\mathbf{C}_{n}^{R}(5)\right)+\nabla_{\tau}\left(\mathbf{C}_{\tau}^{L}(5)+\mathbf{C}_{\tau}^{R}(5)\right)\right] \\
&-3 \bar{U}_{\tau}\left(m_{22}+2 \bar{U}_{\tau} \mathbf{F}_{i j}(3)-\left(\bar{U}_{\tau}\right)^{2} \mathbf{F}_{i j}(1)\right)+3\left(\bar{U}_{\tau}\right)^{2} \mathbf{F}_{i j}(3)-\left(\bar{U}_{\tau}\right)^{3} \mathbf{F}_{i j}(1),
\end{aligned}
\label{eq43}
\end{equation}

\begin{equation}
\begin{aligned}
m_{223} &=\mathbf{C}^{L}(6)+\mathbf{C}^{R}(6)-\Delta t\left[\nabla_{n}\left(\mathbf{C}_{n}^{L}(6)+\mathbf{C}_{n}^{R}(6)\right)+\nabla_{\tau}\left(\mathbf{C}_{\tau}^{L}(6)+\mathbf{C}_{\tau}^{R}(6)\right)\right]\\
&-\bar{U}_{\tau} m_{33}.
\end{aligned}
\label{eq44}
\end{equation}

The formulations of introduced parameters including $\mathbf{C}$, $\mathbf{C}_{n}$ and $\mathbf{C}_{\tau}$ could also be found in \ref{C:1}. 

\subsection{\emph{Gas-surface Boundary Condition}}
\label{SS:2-4}

The Boundary condition (BC) plays an essential role in representing the gas-surface interaction near the wall. Based on the kinetic theory, the Maxwell boundary condition could be given by

\begin{equation}
{f}_{{BC}}=\frac{\rho_{{W}}}{\left(2 \pi {R}_{{g}} {T}_{{W}}\right)^{{D} / 2}} \exp \left[-\frac{\left(\bm{\xi}-{U}_{{W}}\right)^{2}}{2 {R}_{{g}} {T}_{{W}}}\right],
\label{eq45}
\end{equation}

\noindent where $\rho_{{W}}, \mathbf{U}_{{W}}$ and ${T}_{{W}}$ denote the wall density, the wall velocity and the wall temperature, respectively. The Maxwell boundary condition assumes that the gas molecule would be reflected diffusely. Usually, the wall velocity $\mathbf{U}_{{W}}$ and the wall temperature ${T}_{{W}}$ could be determined by the wall condition. The wall density $\rho_{{W}}$ is calculated based on the density colliding with the wall. Assuming that the wall is on the left side of the interface, the wall density $\rho_{{W}}$ can be computed as

\begin{equation}
\rho_{W}=\frac{\left(2 \pi R_{g} T_{W}\right)^{D / 2}\left\langle\bm{\xi} f_{i j}\left(\mathbf{x}_{i j}, \bm{\xi}, t\right)\right\rangle_{<0}}{\left\langle\bm{\xi} \exp \left[-\frac{\left(\bm{\xi}-\mathbf{U}_{W}\right)^{2}}{2 R_{g} T_{W}}\right]\right\rangle_{>0}}.
\label{eq46}
\end{equation}

Then, the VDF at the wall interface can be fully determined and the discrete form of numerical flux could be reconstructed by the numerical integration as Eq. (\ref{eq21}-\ref{eq23}). For the explicit form of G13-MGKS, the wall density $\rho_{{W}}$ takes the form of \cite{liu_development_2022}

\begin{equation}
\rho_{W}=-\frac{\mathbf{A}^{R}(2)-\Delta t\left[\nabla_{n} \mathbf{A}_{n}^{R}(2)+\nabla_{\tau} \mathbf{A}_{\tau}^{R}(2)\right]}{\left(U_{W} / 2\right)\left[1+\operatorname{erf}\left(\sqrt{\lambda_{W}} U_{W}\right)\right]+\left(2 \sqrt{\lambda_{W} \pi}\right)^{-1} \exp \left(-\lambda_{W} U_{W}^{2}\right)},
\label{eq47}
\end{equation}

\noindent where $\lambda_{w}=1 / 2 R_{g} T_{w}$. Now all the required macroscopic values are given and the integration parameters can be calculated following the similar procedure introduced in Section \ref{SS:2-2}. The only difference is that the moment of VDF $\left\langle\xi_{n}^{o} C_{\tau}^{p} \zeta^{q} f_{i j}\right\rangle_{>0}$ could be simplified from Eq. (\ref{eqA2}) to

\begin{equation}
\left\langle\xi_{n}^{o} C_{\tau}^{p} \zeta^{q} f_{i j}\right\rangle_{>0}=\left\langle\xi_{n}^{o} C_{n}^{0}\right\rangle_{>0}^{eq}\left\langle C_{\tau}^{p}\right\rangle^{eq}\left\langle\zeta^{q}\right\rangle^{eq}.
\label{eq48}
\end{equation}

\subsection{\emph{Computational sequence}}
\label{SS:2-5}

\noindent For the discrete form of G13-MGKS:
\begin{itemize}
\setlength{\itemsep}{0pt}
\setlength{\parsep}{0pt}
\setlength{\parskip}{0pt}
\item[1)]
Determine the time step based on the CFL condition. 
\item[2)]
Calculate derivatives of macroscopic quantities, stress and heat flux by the van Leer limiter. Interpolate the macroscopic quantities, stress and heat flux to the surrounding points around the cell interface $\mathbf{x}_{s}=\mathbf{x}_{i j}-\bm{\xi}\Delta t$ by Eq. (\ref{eq15}).
\item[3)]
Construct the VDF at the cell interface $f_{ij}\left(\mathbf{x}_{i j}, \bm{\xi}, t^{n+1}\right)$ by Eqs. (\ref{eq16}). Calculate the numerical fluxes related to macroscopic variables $\mathbf{F}_{i j}$, stress $\mathbf{G}_{i j}$ and heat flux $\mathbf{H}_{i j}$ by Eqs. (\ref{eq21})-(\ref{eq23}), respectively.
\item[4)]
Update the macroscopic governing equations for the macroscopic variables $\mathbf{W}_{i}^{n+1}$, stress $\boldsymbol{\sigma}_{i}^{n+1}$ and heat flux $\mathbf{q}_{i}^{n+1}$ by Eq. (\ref{eq10}), Eq. (\ref{eq17}) and Eq. (\ref{eq18}), respectively. 
\item[5)]
Repeat steps (1)–(4) before the convergence criterion can be satisfied.
\end{itemize}

\noindent For the explicit form of G13-MGKS:
\begin{itemize}
\setlength{\itemsep}{0pt}
\setlength{\parsep}{0pt}
\setlength{\parskip}{0pt}
\item[1)]
Determine the time step based on the CFL condition. 
\item[2)]
Calculate integration parameters $\mathbf{A}$, $\mathbf{A}_{n}$ and $\mathbf{A}_{\tau}$, $\mathbf{B}$, $\mathbf{B}_{n}$ and $\mathbf{B}_{\tau}$, $\mathbf{C}$, $\mathbf{C}_{n}$ and $\mathbf{C}_{\tau}$ by Eqs. (\ref{eq25})-(\ref{eq28}) and Eqs. (\ref{eqC1})-(\ref{eqC10}). The derivatives of these integration parameters could be obtained by the van Leer limiter.
\item[3)]
Calculate the numerical fluxes related to macroscopic variables $\mathbf{F}_{i j}$, stress $\mathbf{G}_{i j}$ and heat flux $\mathbf{H}_{i j}$ by Eq. (\ref{eq24}), Eqs. (\ref{eq33})-(\ref{eq36}) and Eqs. (\ref{eq37})-(\ref{eq38}), respectively.
\item[4)]
Update the macroscopic governing equations for the macroscopic variables $\mathbf{W}_{i}^{n+1}$, stress $\boldsymbol{\sigma}_{i}^{n+1}$ and heat flux $\mathbf{q}_{i}^{n+1}$ by Eq. (\ref{eq10}), Eq. (\ref{eq17}) and Eq. (\ref{eq18}), respectively. 
\item[5)]
Repeat steps (1)–(4) before the convergence criterion can be satisfied.
\end{itemize}

\section{NUMERICAL EXPERIMENTS}
\label{S:3}
In present section, several numerical cases are adopted to verify the present method for rarefied flows, including the shock wave structure, the unsteady Sod shock tube, the lid-driven cavity flow and the unsteady Rayleigh flow. The hard-sphere argon gas is considered. The Prandtl number is given as $\operatorname{Pr}=1$ and specific heat ratio is given as $\gamma=5/3$ in the present section. For ease of description and labeling, the results of the discrete form and explicit form of the present method are denoted as “Discrete” and “Explicit,” respectively.

\subsection{\emph{The Shock Wave Structure}}
\label{sec3-1}

The first numerical case tested in the present section is the shock wave structure. At the start of the simulation, the left and right boundaries are given by the Rankine-Hugoniot condition \cite{rankine_xv_1870} as

\begin{equation}
\begin{gathered}
M_{R}=\sqrt{\frac{(\gamma-1) M_{L}^{2}+2}{2 \gamma M_{L}^{2}-(\gamma-1)}}, \\
\frac{\rho_{R}}{\rho_{L}}=\frac{(\gamma+1) M_{L}^{2}}{(\gamma-1) M_{L}^{2}+2}, \\
\frac{T_{R}}{T_{L}}=\frac{\left(1+\frac{\gamma-1}{2} M_{L}^{2}\right)\left(\frac{2 \gamma}{\gamma-1} M_{L}^{2}-1\right)}{\left(\frac{2 \gamma}{\gamma-1}+\frac{\gamma-1}{2}\right) M_{L}^{2}}.
\end{gathered}
\label{eq49}
\end{equation}

\noindent where $\rho$, $M$ and $T$ denote the density, Mach number and temperature. The subscripts $L$ and $R$ represent values in upstream and downstream of the flow field, respectively. The reference Mach number Ma is specified as the upstream Mach number $M_{L}$. The viscous could be calculated by 

\begin{equation}
\mu=\mu_{r e f}\left(\frac{T}{T_{0}}\right)^{\omega},
\label{eq50}
\end{equation}

\noindent where the temperature dependency index is adopted as $\omega=0.81$. The variable hard sphere (VHS) model \cite{huang_unified_2013} could be used for the reference viscosity coefficient as

\begin{equation}
\mu_{r e f}=\frac{15 \sqrt{\pi}}{2(5-2 \omega)(7-2 \omega)} \mathrm{Kn}.
\label{eq51}
\end{equation}

Here, the reference Knudsen number $\mathrm{Kn}$ is set as 1.0, which means that the reference length is equal to the mean free path in the upstream of flow field. The time step can be determined based on a CFL number $\sigma_{\mathrm{CFL}}$ equal to 0.95. Moreover, 100 mesh points are utilized to discrete the physical domain in the range of $x \in[-25,25]$. For the discrete form of G13-MGKS, the Newton-Cotes quadrature is uniformly utilized with 151 discrete points distributed in $\left[-15\sqrt{2 R T_{0}}, 15\sqrt{2 R T_{0}}\right]$.

\begin{figure}[H]
\centering
\includegraphics[width=7cm]{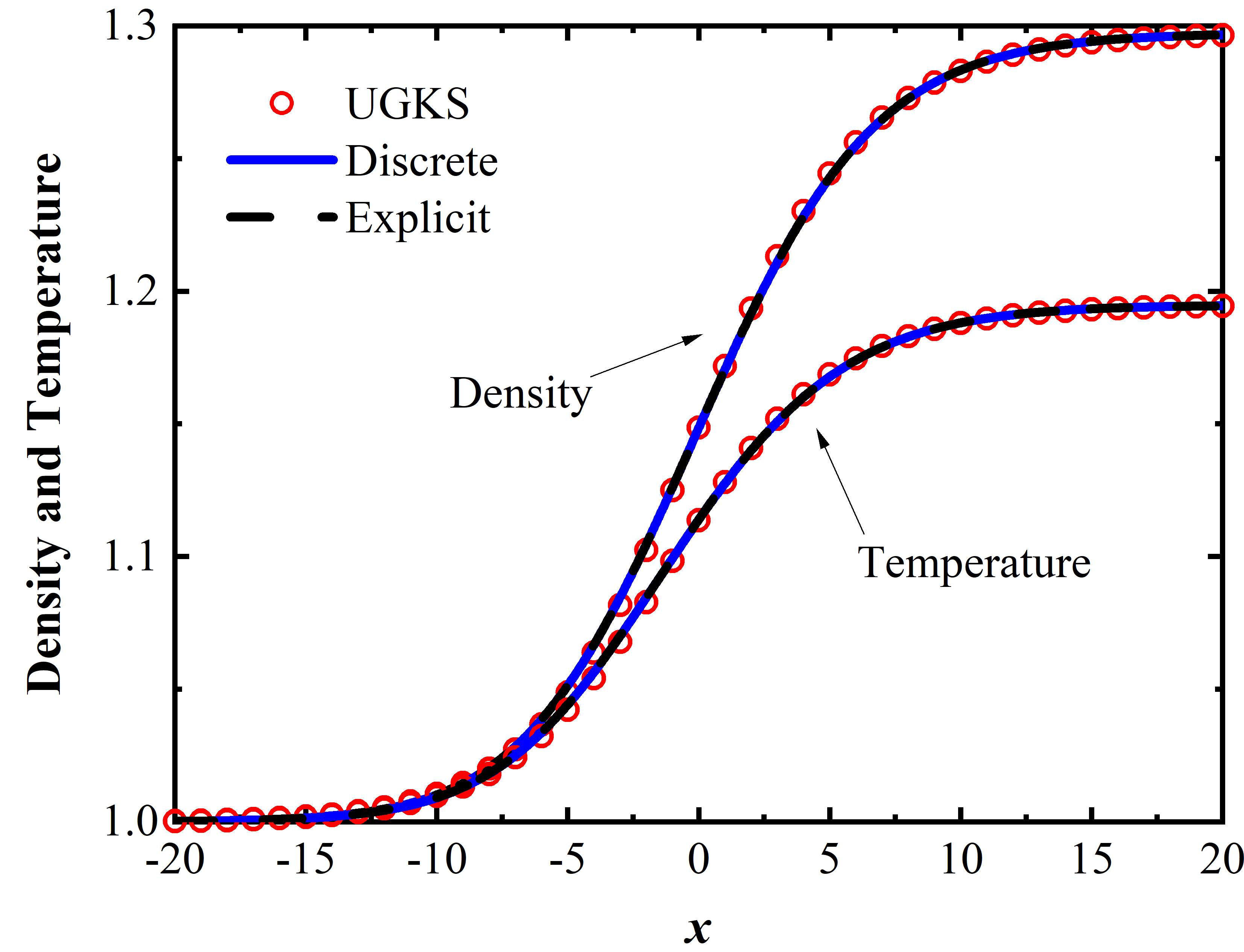}
\includegraphics[width=7cm]{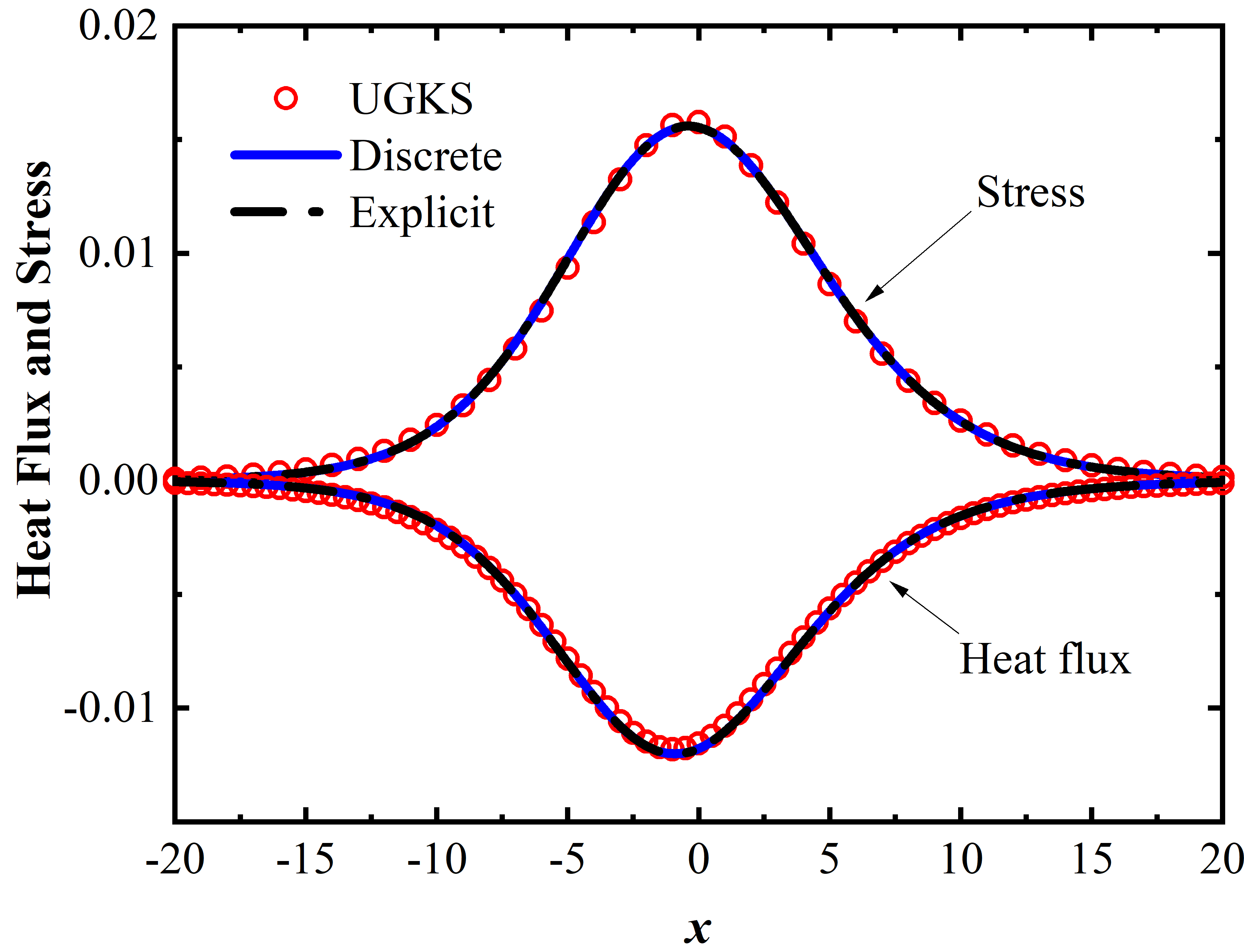}
\caption{Shock structure at Ma = 1.2 (maximum $\mathrm{Kn}_{L}$ is about 0.023): (Left) density and temperature, (right) stress and heat flux.}
\label{Fig1}
\end{figure}

\begin{figure}[H]
\centering
\includegraphics[width=7cm]{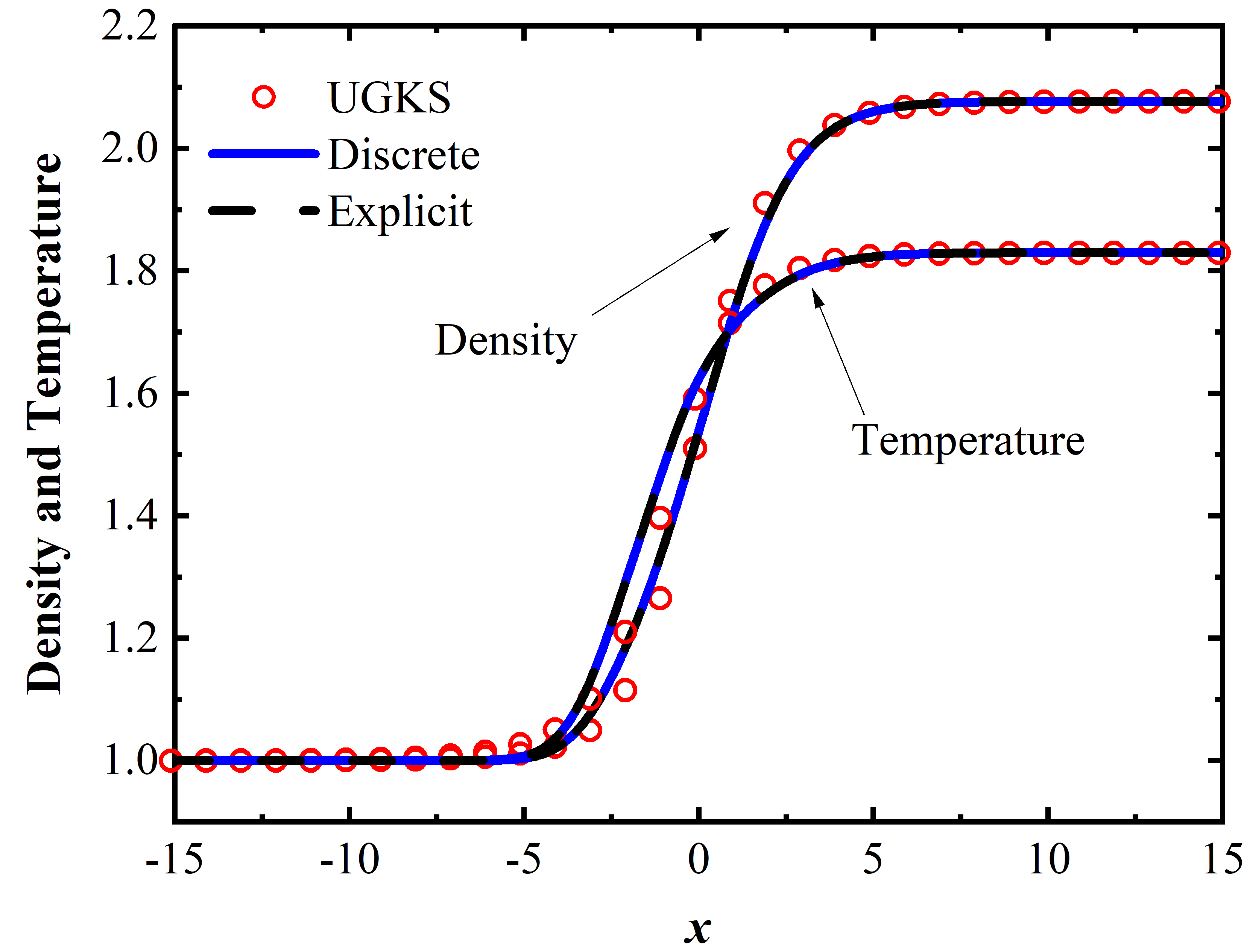}
\includegraphics[width=7cm]{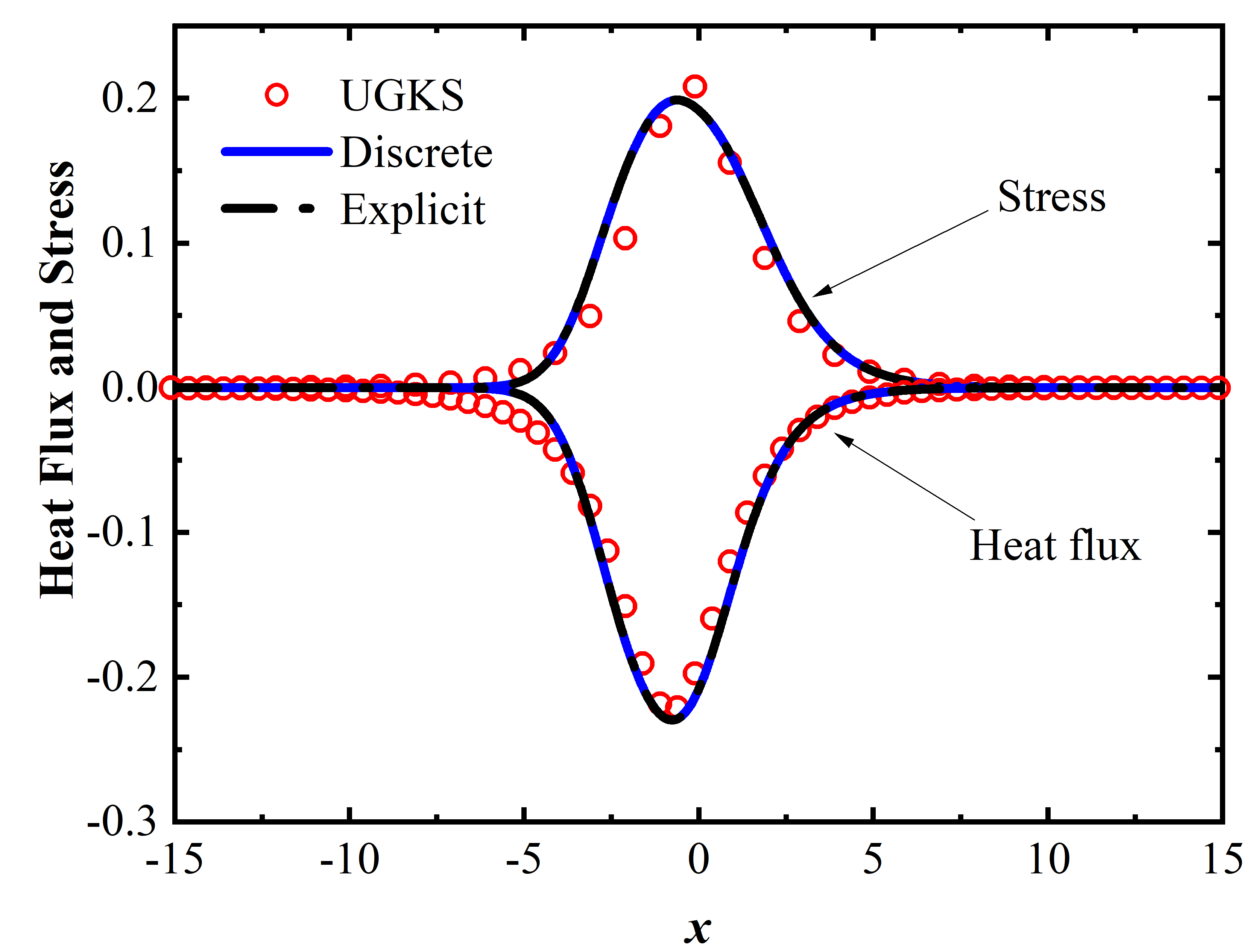}
\caption{Shock structure at Ma = 1.8 (maximum $\mathrm{Kn}_{L}$ is about 0.206): (Left) density and temperature, (right) stress and heat flux.}
\label{Fig2}
\end{figure}

Since the UGKS has been extensively validated in earlier publications \cite{xu_unified_2010, huang_unified_2012, chen_three-dimensional_2020}, the G13-MGKS results can be compared to the references from the UGKS and use the same spatial meshes. All the solutions are output at the time of $t=250$. The profiles of density, temperature, stress and heat flux when Ma=1.2 are shown in Fig. \ref{Fig1}. It can be found the solutions from the present solver could match well with the references. To represent the local rarefied effect in the shock wave, the local Knudsen number $\mathrm{Kn}_{L}$ is given by

\begin{equation}
\mathrm{Kn}_{L}=\lambda_{L} \max \left(\frac{1}{\rho} \nabla_{\mathbf{x}} \rho, \frac{1}{T} \nabla_{\mathbf{x}} T\right),
\label{eq52}
\end{equation}

\noindent where the local mean free path equals $\lambda_{L}=16 \mu / 5 \sqrt{2 \rho T}$. Based on Eq. (\ref{eq52}), the maximum $\mathrm{Kn}_{L}$ is about 0.023 when the Ma =1.2. Further increasing the Mach number to 1.8 and the maximum $\mathrm{Kn}_{L}$ rises to about 0.206, deviations appear in Fig. \ref{Fig2} compared to the reference data. Basically, reasonable results can still be described by the G13 distribution function. 

\begin{figure}[H]
\centering
\includegraphics[width=7cm]{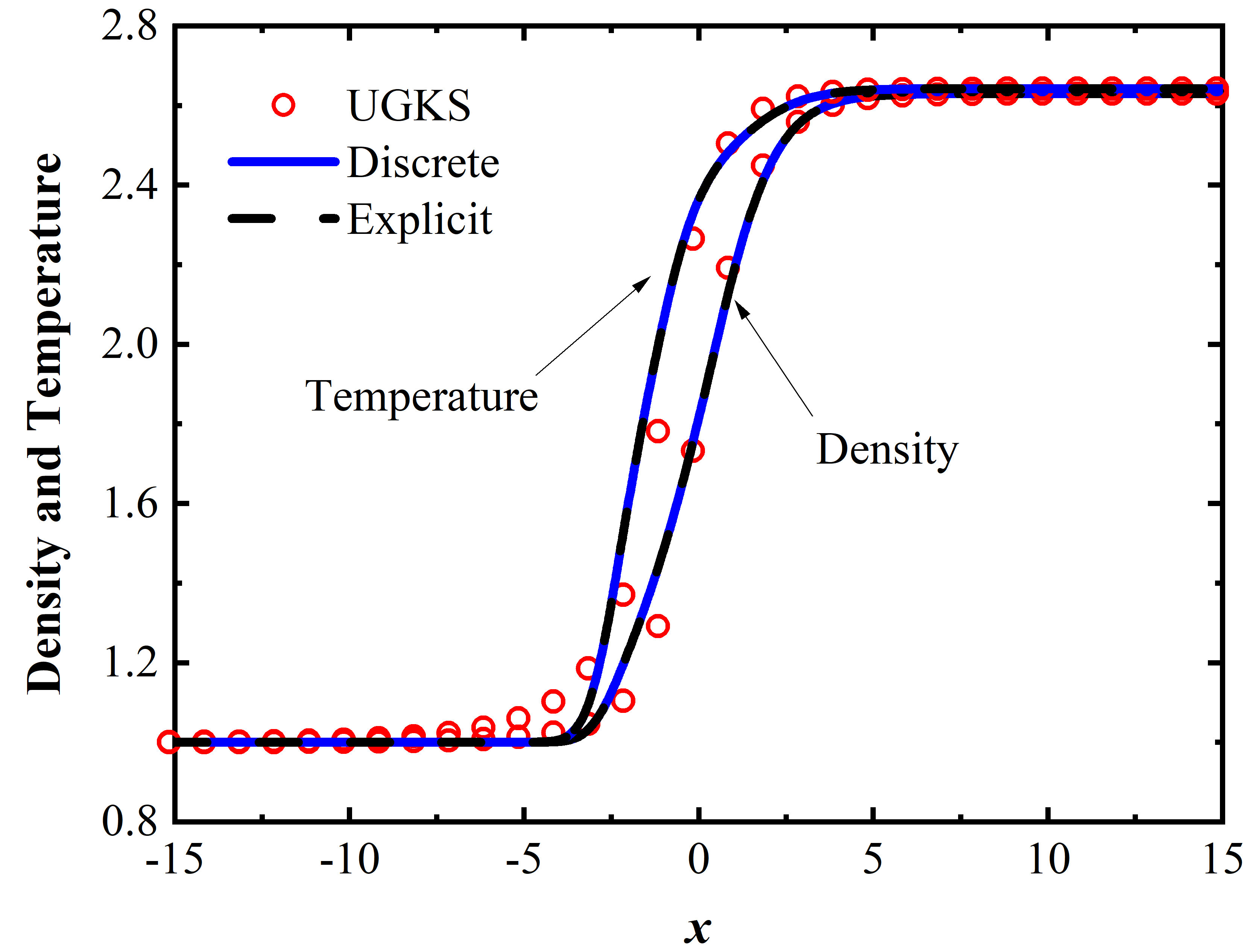}
\includegraphics[width=7cm]{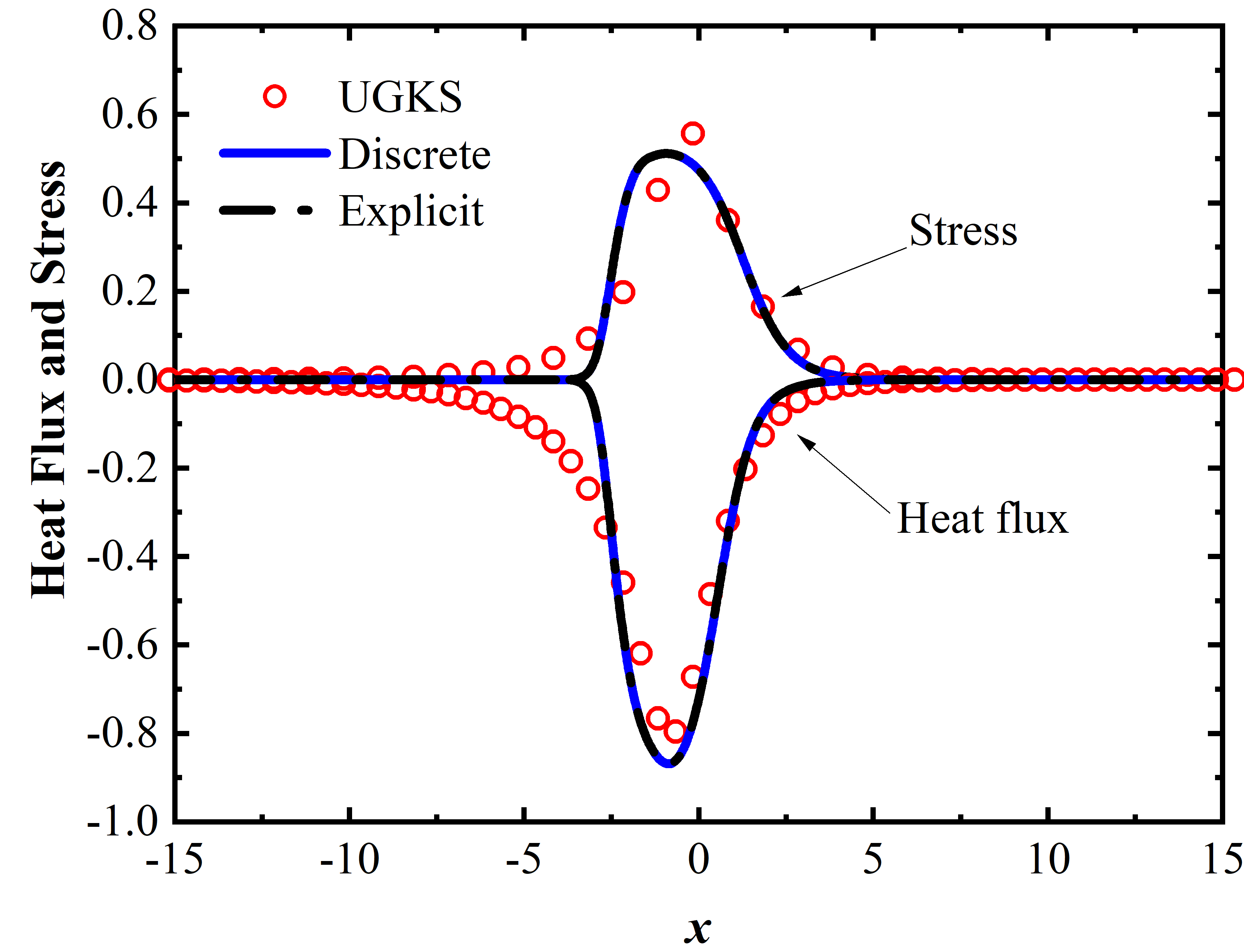}
\caption{Shock structure at Ma = 2.4 (maximum $\mathrm{Kn}_{L}$ is about 0.402): (Left) density and temperature, (right) stress and heat flux.}
\label{Fig3}
\end{figure}

As the Mach number increases to 2.4 and the maximum $\mathrm{Kn}_{L}$ increases to 0.402, obvious deviations can be found in Fig. \ref{Fig3}, especially at the position of $x=\pm 5.0$ between the the subsonic flow with high temperature and supersonic flow with low temperature, substantial density and temperature gradients develop. The rarefied effect can become significant because of the insufficient collisions of gas molecules in the shock wave.  

\begin{figure}[H]
\centering
\includegraphics[width=7cm]{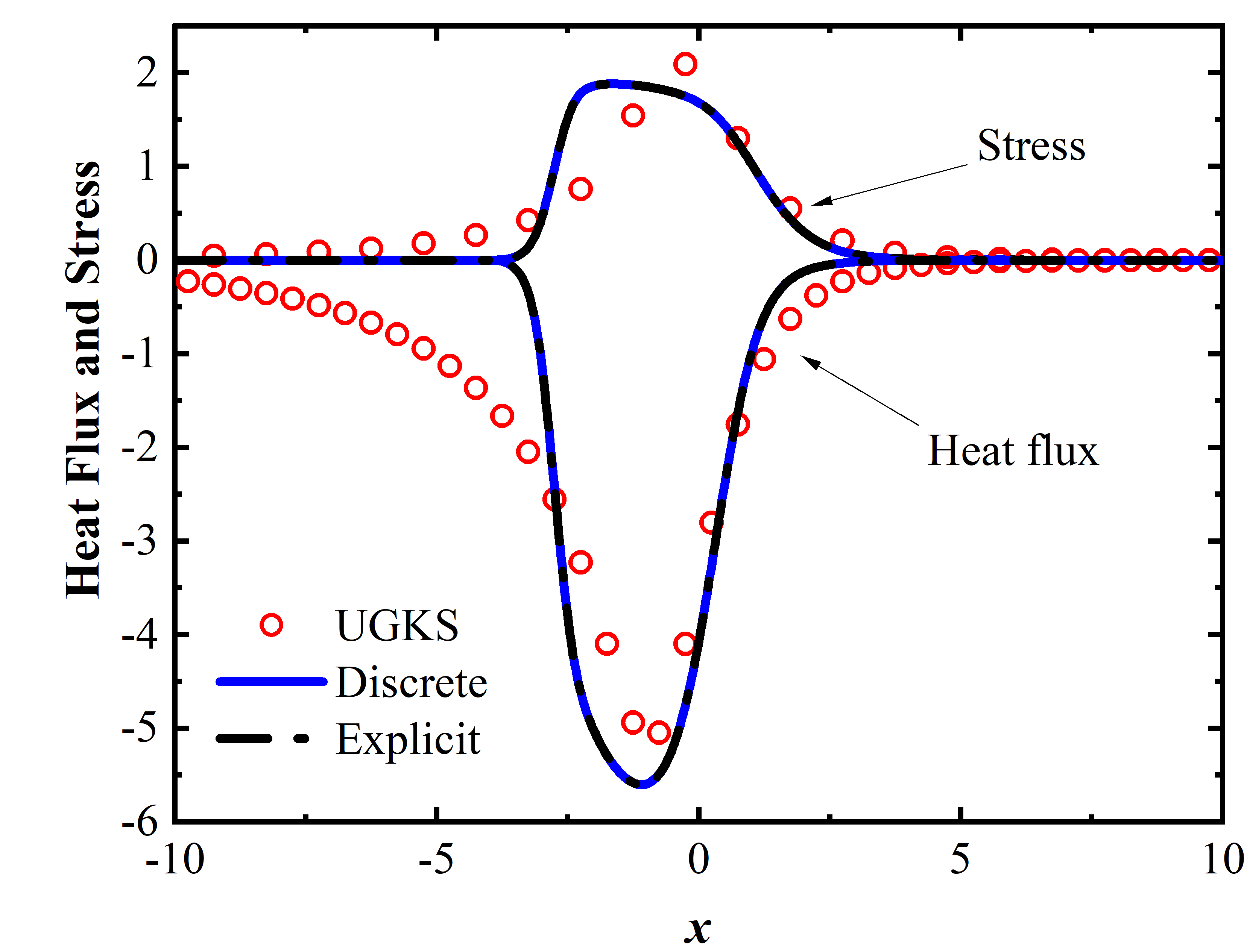}
\includegraphics[width=7cm]{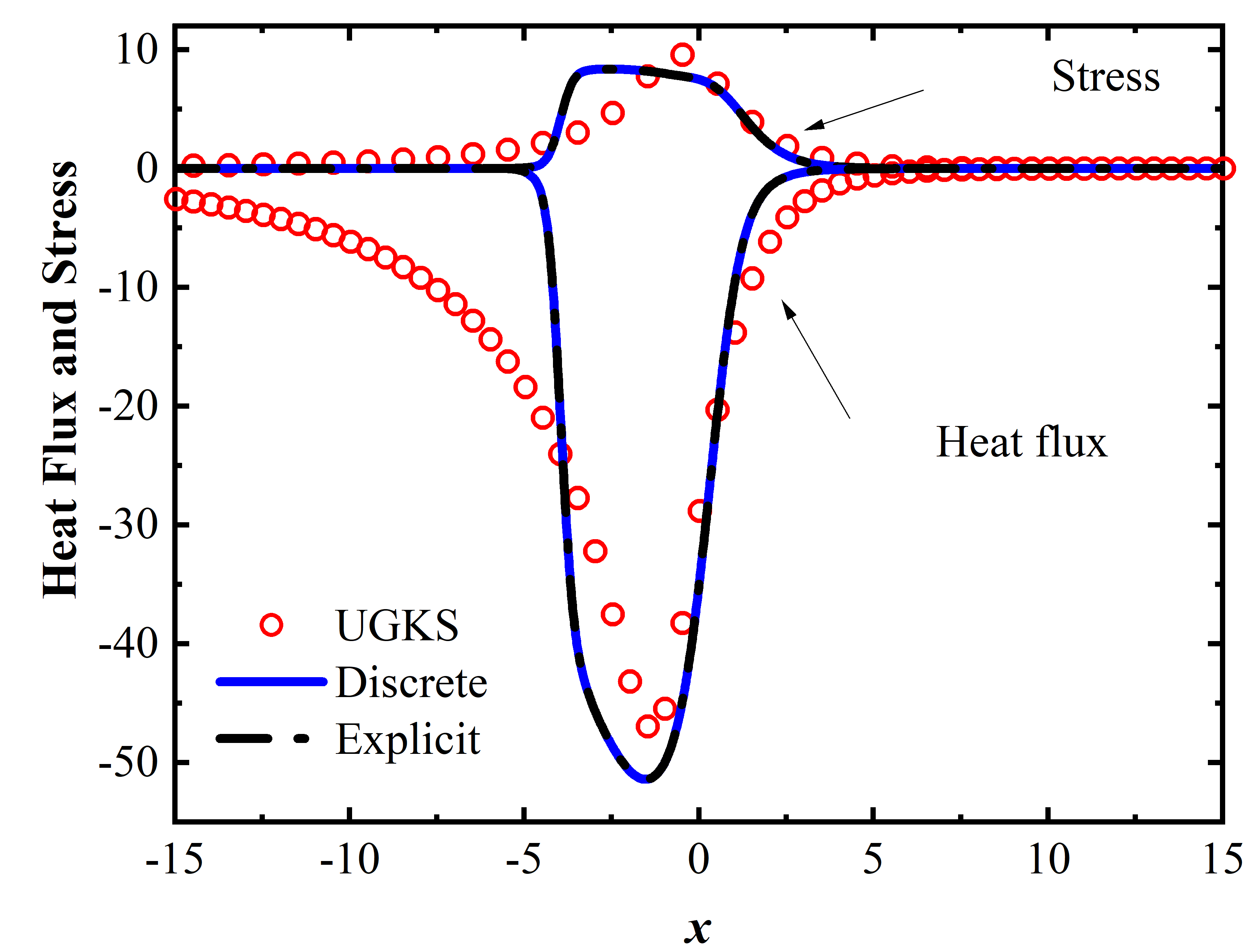}
\caption{Stress and heat flux of Shock structure at (Left) Ma = 4.0 and (right) Ma = 8.0.}
\label{Fig4}
\end{figure}

To verify the stability of the current framework, we tested the results for Ma=4.0 and Ma=8.0 as shown in Fig. \ref{Fig4}. At such conditions of high Mach numbers and strong rarefied effect, the distribution of stress and heat flow in strong shock waves. The sharp differences indicate that G13 as the lowest order distribution function expansion in the Grad series is no longer able to accurately capture the rarefied effects. Higher-order truncated VDF, including but not limited to G26 and G45, should be adopted to obtain better results. In addition, the discrete and explicit versions of the algorithm are essentially identical in terms of results, except for the difference in computational efficiency.   

\subsection{\emph{The Sod Shock Tube}}
\label{sec3-2}

The classical Sod shock tube test is examined at different Knudsen numbers to verify the performance of G13-MGKS in unsteady flow. The gases at two various states are separated in the shock tube at the start of the computation, and their initial dimensionless values are given as

\begin{equation}
(\rho, U, p)= \begin{cases}(0.125,0,0.1), x>0.5 \\ (1,0,1), x \leq 0.5.\end{cases}
\label{eq53}
\end{equation}

\begin{figure}[H]
\centering
\includegraphics[width=7cm]{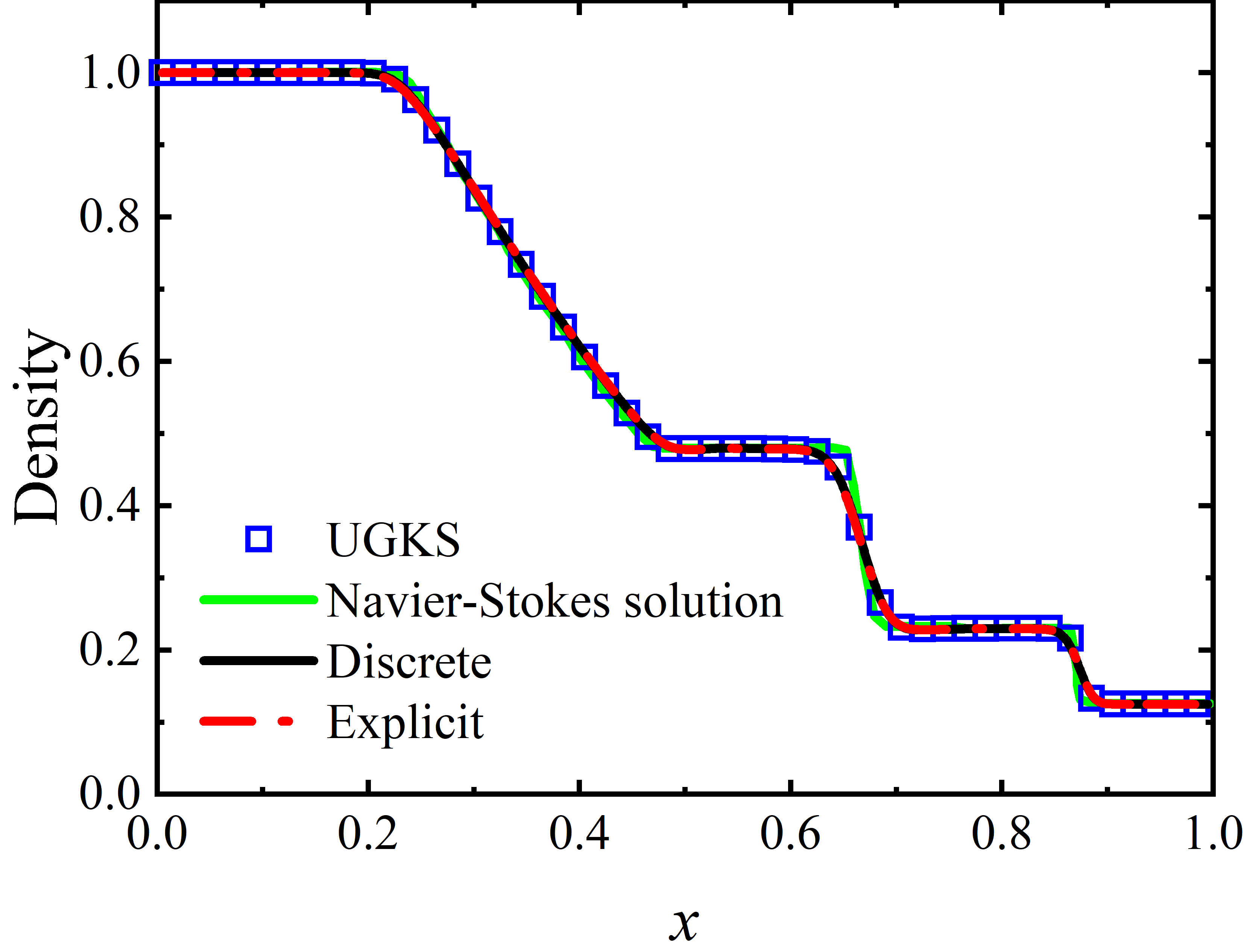}
\includegraphics[width=7cm]{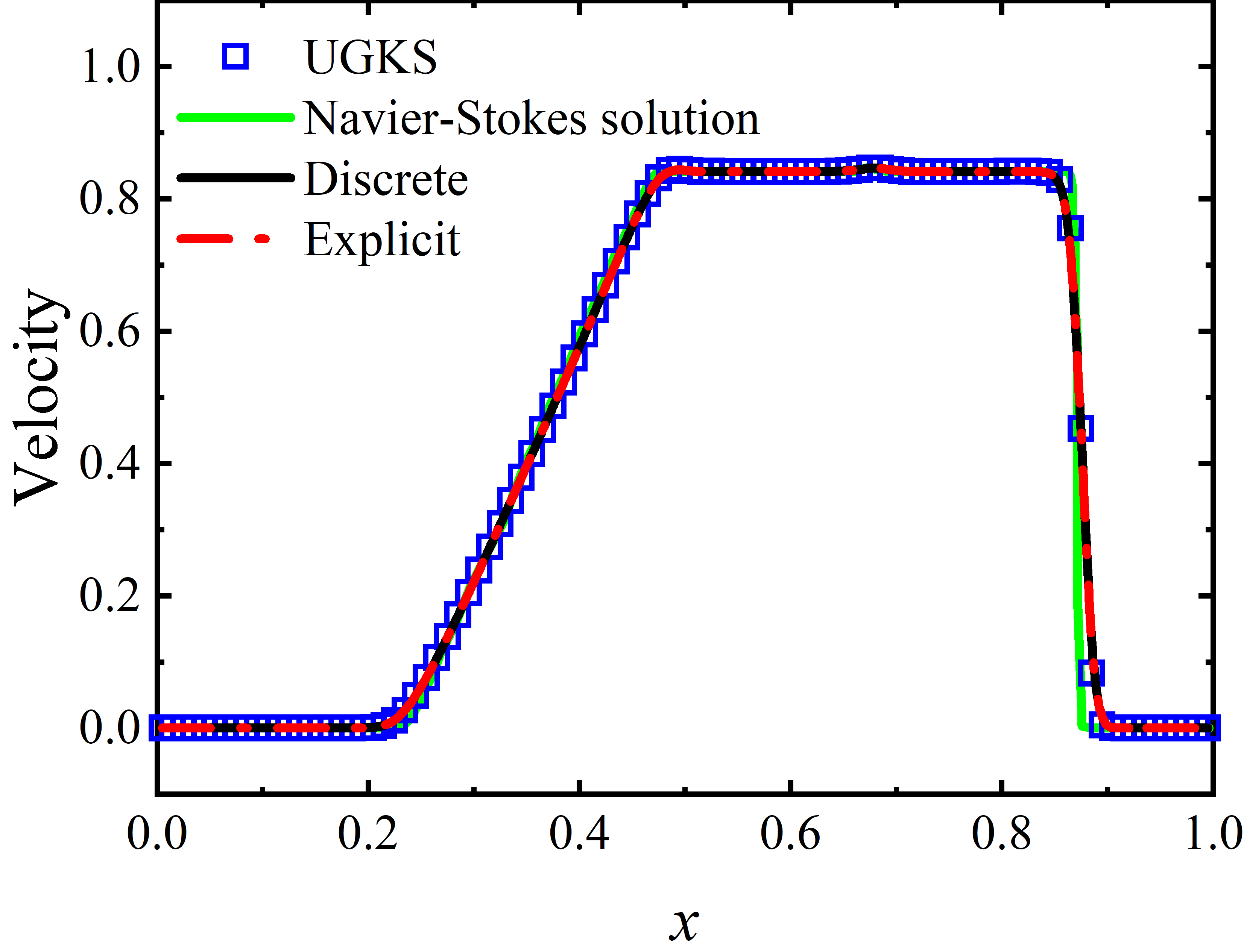}
\caption{Sod shock with reference Knudsen number as $\mathrm{Kn}=1 \times 10^{-4}$, (Left) density and (right) velocity.}
\label{Fig5}
\end{figure}

The viscosity could be also calculated by Eq. (\ref{eq50}) and Eq. (\ref{eq51}) with $\omega=0.81$. The CFL number $\sigma_{\mathrm{CFL}}$ is set as 0.95. To verify the discrete and explicit form of the present method, the references from the Navier-Stokes equations and the UGKS are given for the comparisons when $\mathrm{Kn}=1 \times 10^{-4}$, $\mathrm{Kn}=1 \times 10^{-3}$, $\mathrm{Kn}=1 \times 10^{-2}$  and $\mathrm{Kn}=1 \times 10^{-1}$. All of the solutions are based on calculations done at $t=0.2$. As presented in Fig. \ref{Fig5}, the profiles of density and velocity obtained from Navier-Stokes equations and UGKS are basically identical, except that minor differences appear at the location of rarefaction wave and contact discontinuity. The solutions of discrete and explicit form at $\mathrm{Kn}=1 \times 10^{-4}$ and $t$ = 0.2 could accurately match the benchmark solutions from Navier-Stokes equations and UGKS.  
 
\begin{figure}[H]
\centering
\includegraphics[width=7cm]{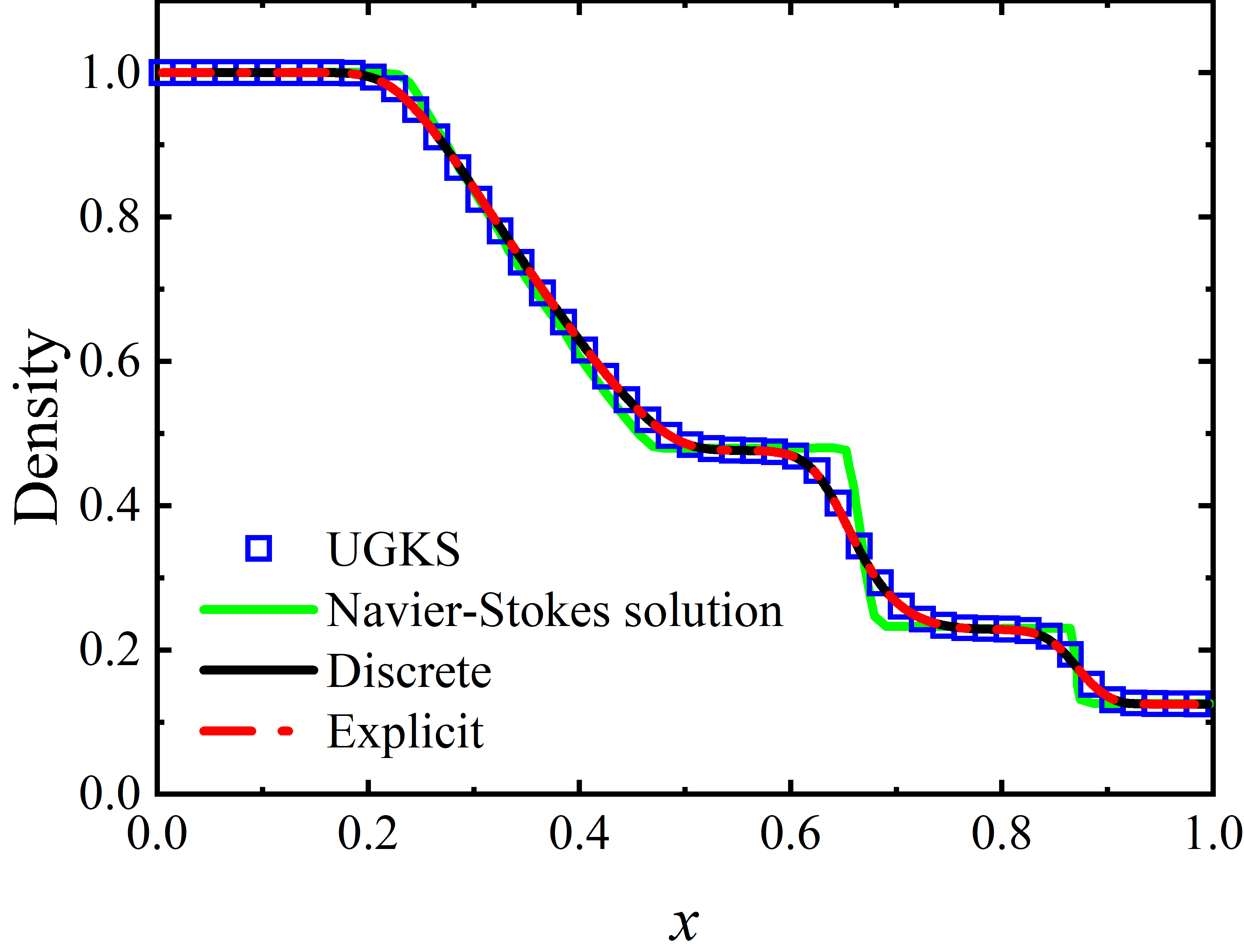}
\includegraphics[width=7cm]{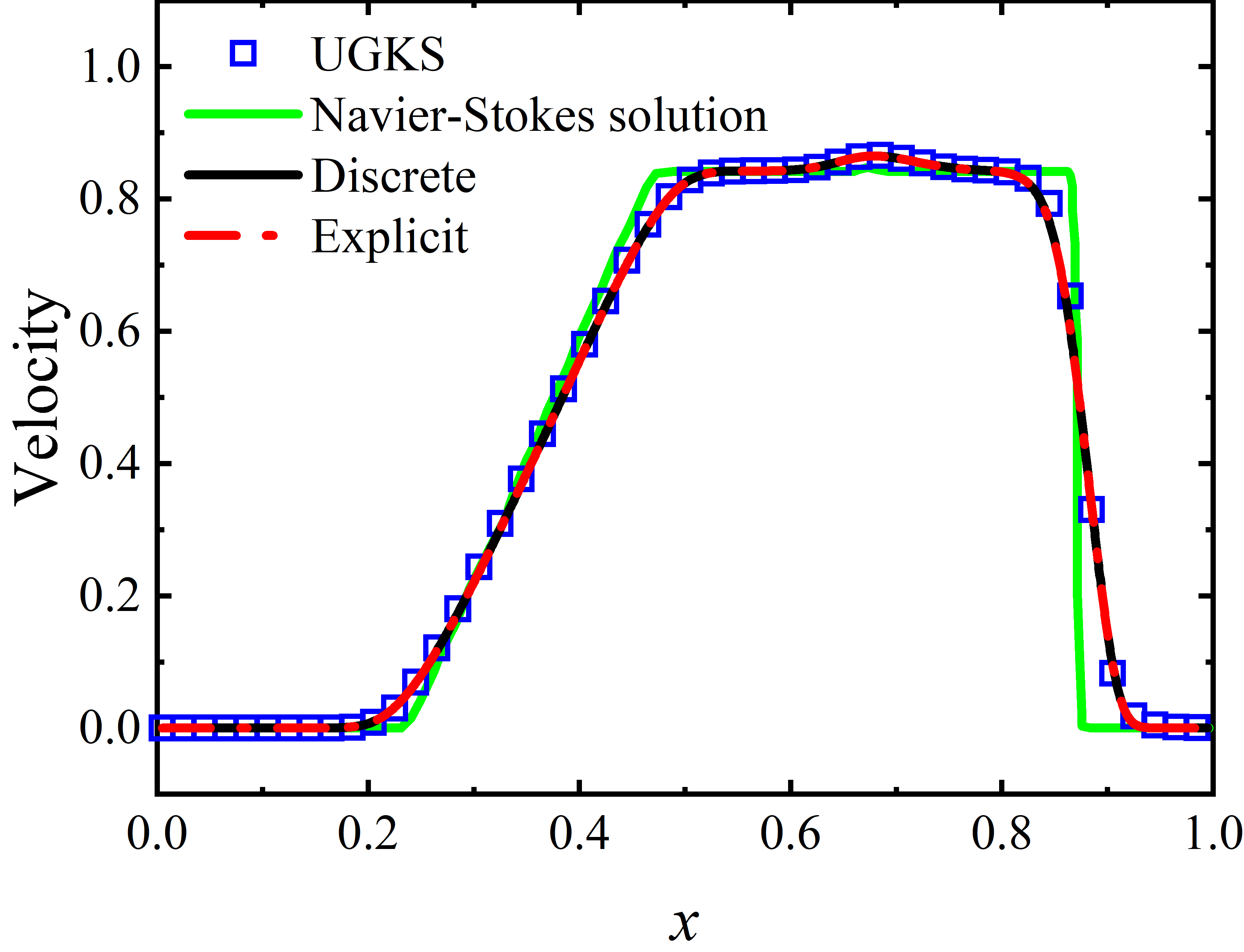}
\caption{Sod shock with reference Knudsen number as $\mathrm{Kn}=1 \times 10^{-3}$, (Left) density and (right) velocity.}
\label{Fig6}
\end{figure}

\begin{figure}[H]
\centering
\includegraphics[width=7cm]{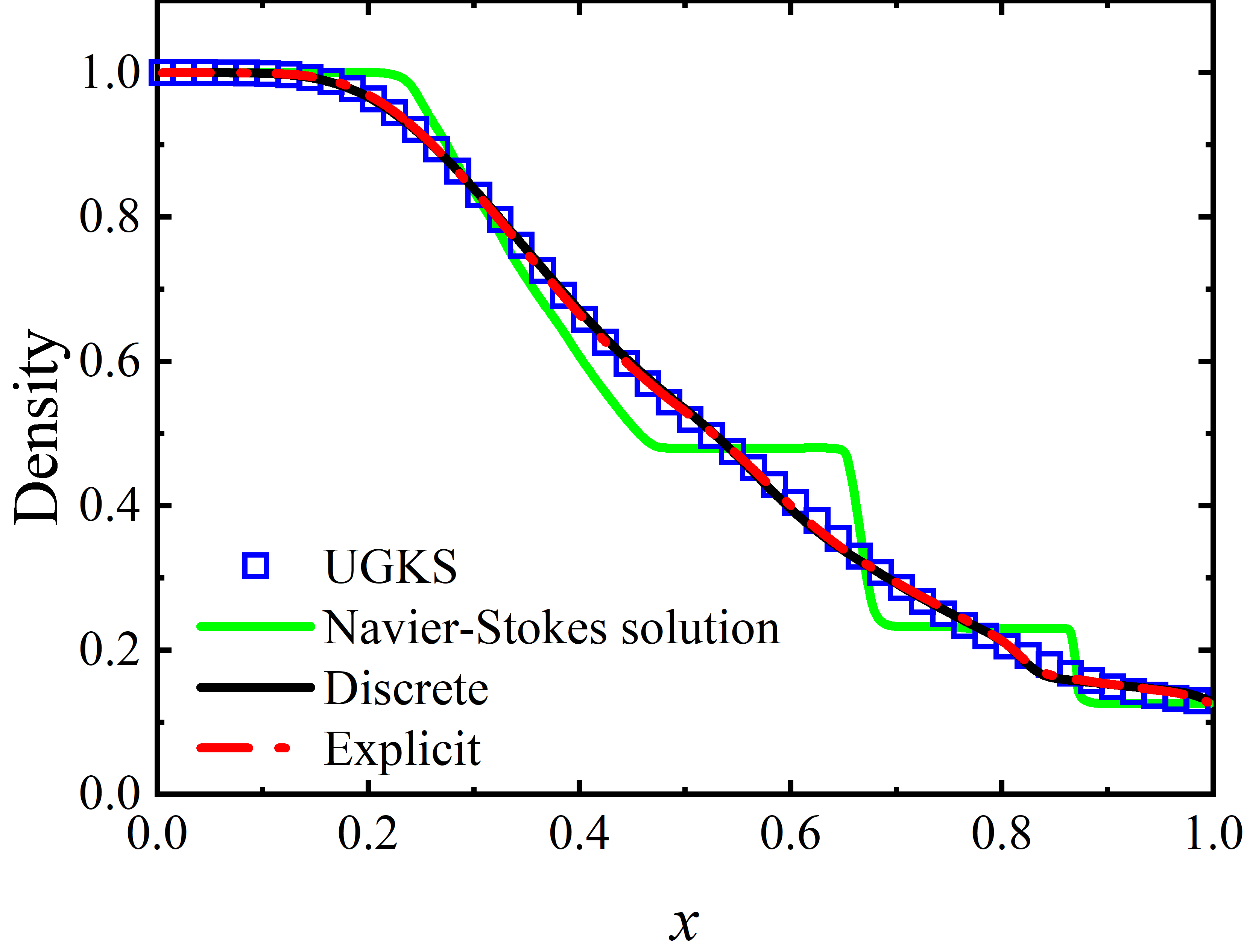}
\includegraphics[width=7cm]{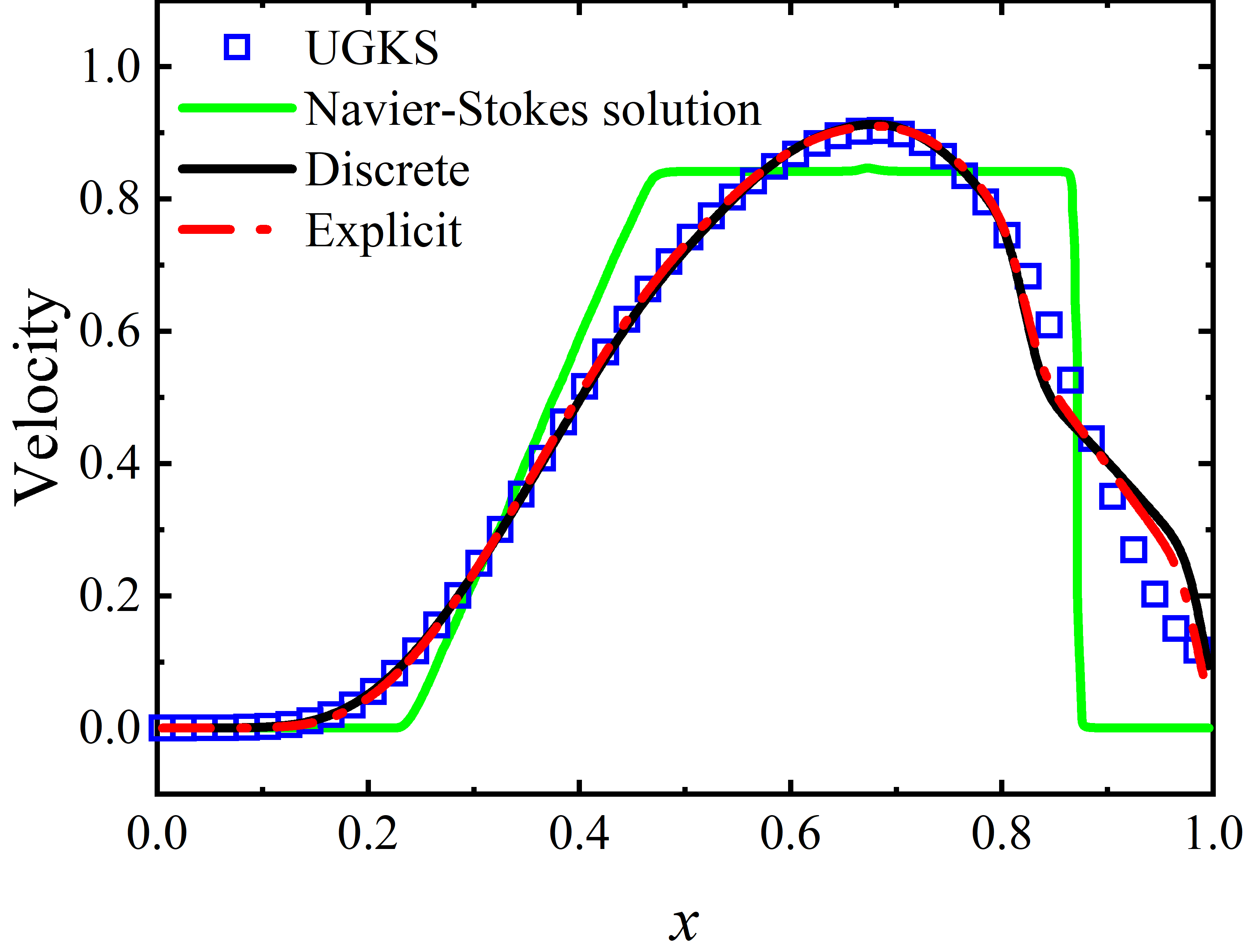}
\caption{Sod shock with reference Knudsen number as $\mathrm{Kn}=1 \times 10^{-2}$, (Left) density and (right) velocity.}
\label{Fig7}
\end{figure}

When the Knudsen number gets to $1 \times 10^{-3}$, the solutions shown in Fig. \ref{Fig6} present smoother profiles. It could be found that the solutions from Navier-Stokes equations deviate from the UGKS solutions apparently while the solutions from present methods recover the UGKS solutions well. As the Knudsen number rises to $1 \times 10^{-2}$, the Navier-Stokes solutions lose their validity and the rarefied effect occupies the whole tube. As presented in Fig. \ref{Fig7}, the numerical solution from the present solver could basically match the UGKS solutions but deviations appear especially for the downstream of velocity. When the $\mathrm{Kn}=1 \times 10^{-1}$ and flow field changes to the transitional regime, the non-equilibrium region enlarges in the downstream. The deviations shown in Fig. \ref{Fig7} indicate that the distribution function in the flow field deviates from the G13 which indicates that a higher-order truncated distribution function is required.    

\begin{figure}[H]
\centering
\includegraphics[width=7cm]{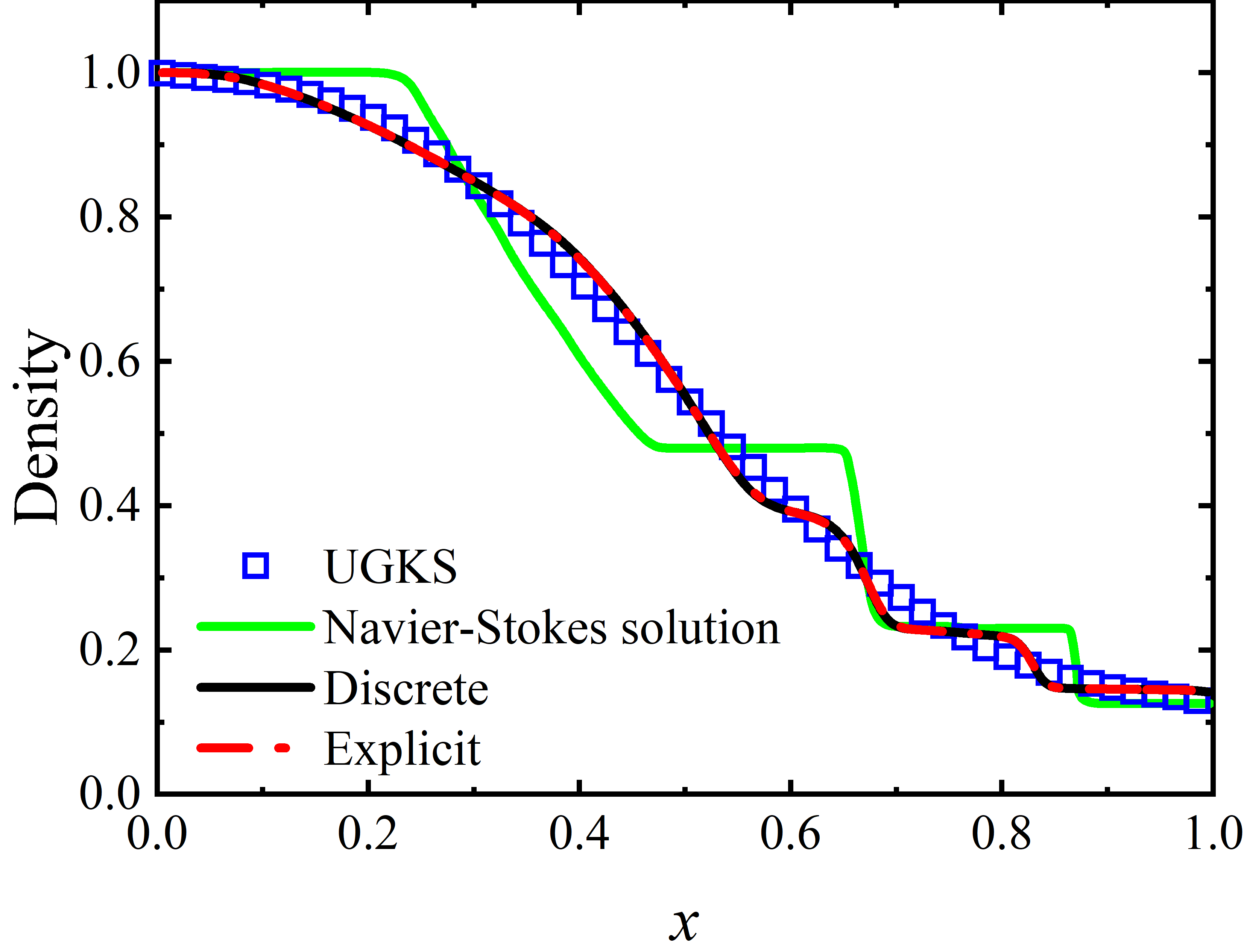}
\includegraphics[width=7cm]{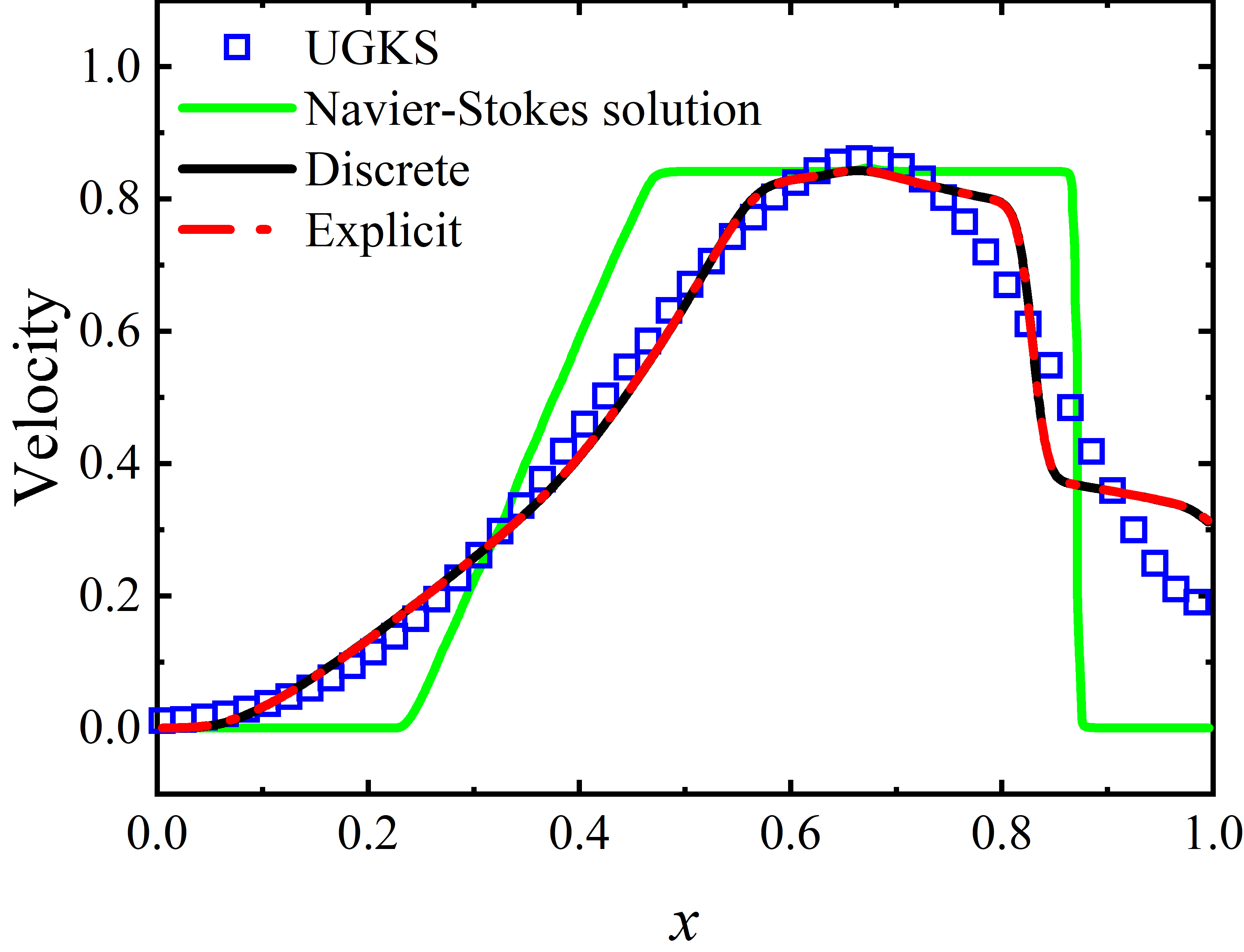}
\caption{Sod shock with reference Knudsen number as $\mathrm{Kn}=1 \times 10^{-1}$, (Left) density and (right) velocity.}
\label{Fig8}
\end{figure}

\subsection{\emph{Lid-driven Cavity Flow}}
\label{sec3-3}

To evaluate the G13-MGKS in the two-dimensional flow, the lid-driven cavity flow is studied at different Knudsen numbers. The square cavity is discretized by $60 \times 60$ uniform mesh points for the computational domain with the edge length of $L=1$. The normalized velocity $U_{W}=0.15$ is fixed at the top boundary for the top lid to drive the flow. The isothermal cavity and walls are set with the normalized temperature $T_{W}=1.0$. The dynamics viscosity could be calculated from Eq. (\ref{eq50}) and Eq. (\ref{eq51}) with the temperature dependency index $\omega=1.0$. The Gauss–Hermite quadrature with $8 \times 8$ velocity points is adopted in the domain of $\left[-4\sqrt{2 R T_{W}}, 4\sqrt{2 R T_{W}}\right]^{2}$ for the discrete form of G13-MGKS. The CFL number is taken as $\sigma_{\mathrm{CFL}}$ = 1.0 and the convergence criteria are determined by the condition that the maximum errors of macroscopic variables between two adjacent iterations are less than $10^{-10}$. 

\begin{figure}[H]
\centering
\includegraphics[width=7cm]{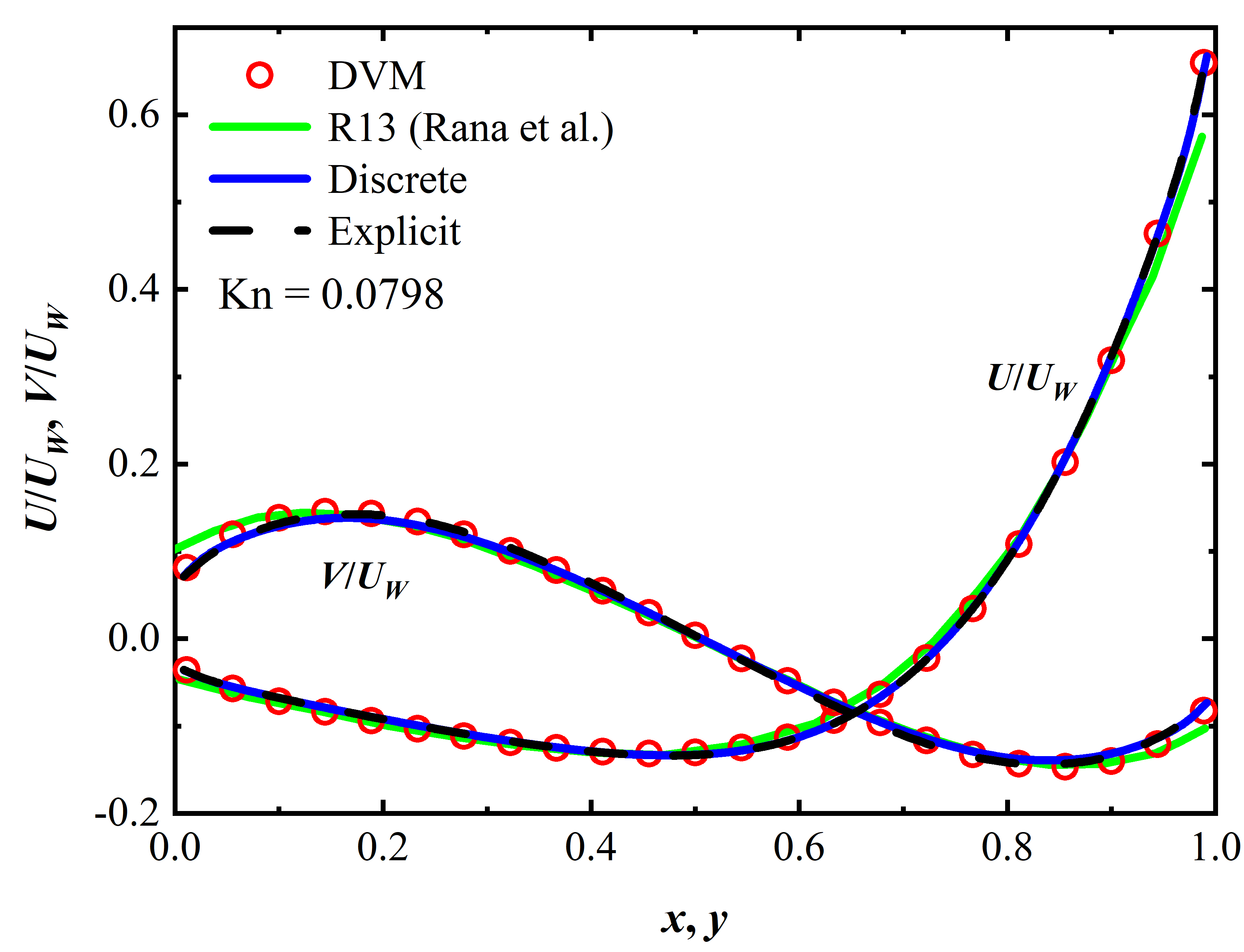}
\includegraphics[width=7cm]{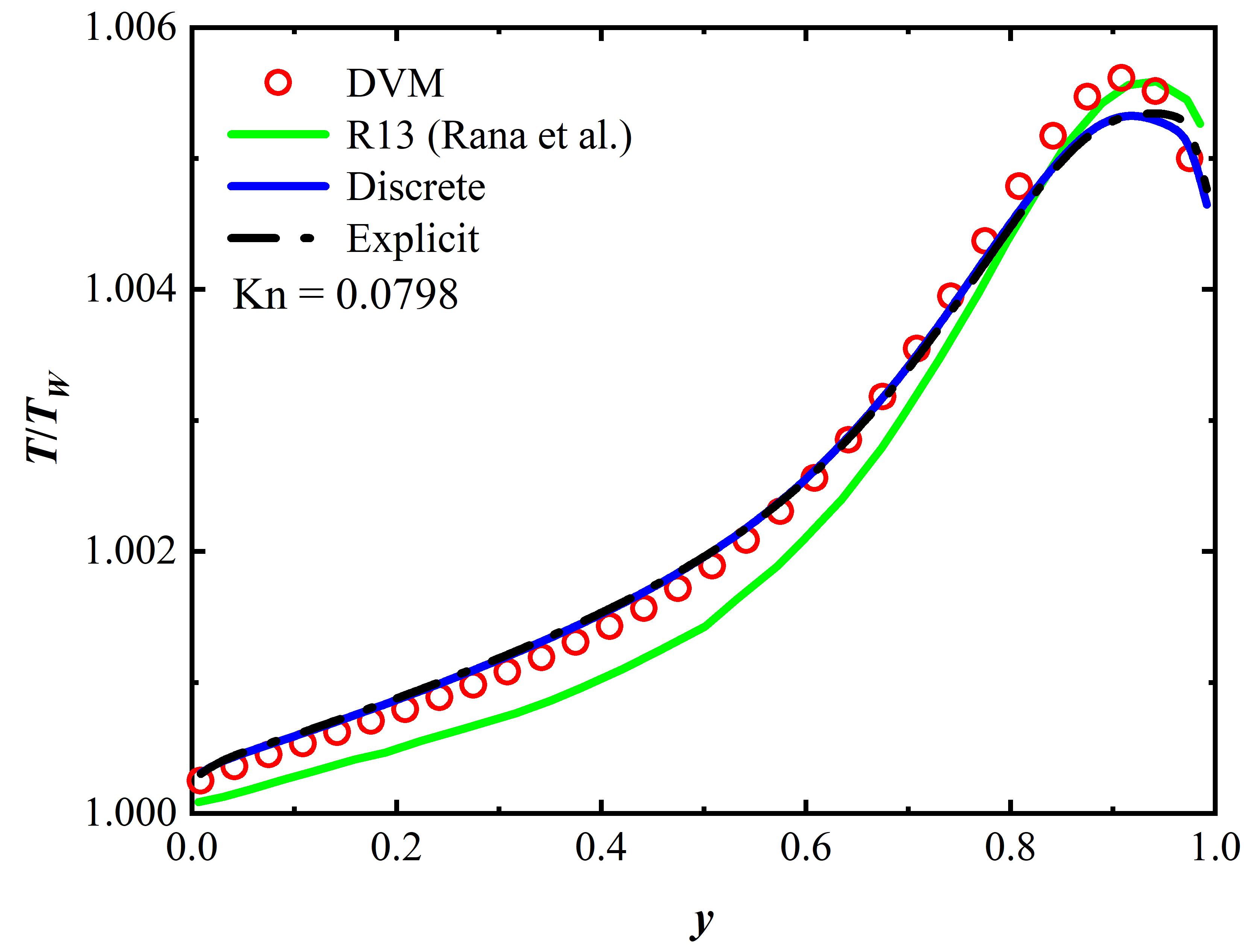}
\caption{Lid-driven cavity flow at $\mathrm{Kn}=0.0798$, (Left) Velocity profiles along the central lines and (right) Temperature profile along the central line.}
\label{Fig9}
\end{figure}

For the validation and comparison, the reference solutions from the DVM and the Moment method with regularized 13-moment equations (R13) \cite{rana_numerical_2015} are given at different Knudsen numbers. The Gauss-Hermite quadrature with $28 \times 28$ mesh points is applied when Kn = 0.0798 and 0.1 while the Gauss-Hermite quadrature with $64 \times 64$ points is utilized when Kn = 0.3989. As shown in Fig. \ref{Fig9}, the profiles of velocity and temperature are presented at Kn = 0.0798. Compared with the solutions from R13, good agreements with the results of DVM can be achieved by both the discrete and explicit forms. The maximum relative errors of the $U/U_{W}$ and $T/T_{W}$ along the horizontal central line are only 0.92\% and 0.81\% at Kn = 0.0798.  

\begin{figure}[H]
\centering
\includegraphics[width=7cm]{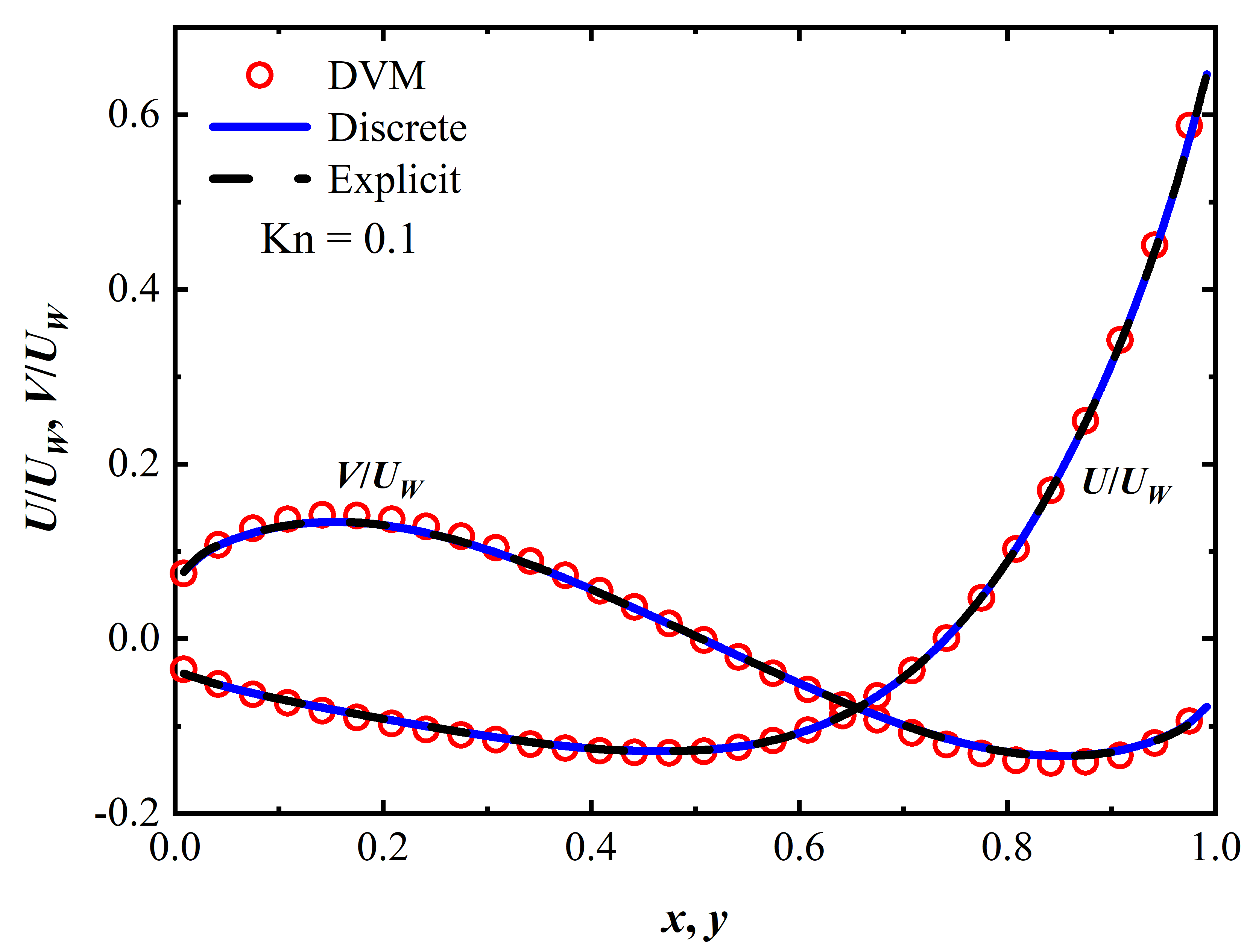}
\includegraphics[width=7cm]{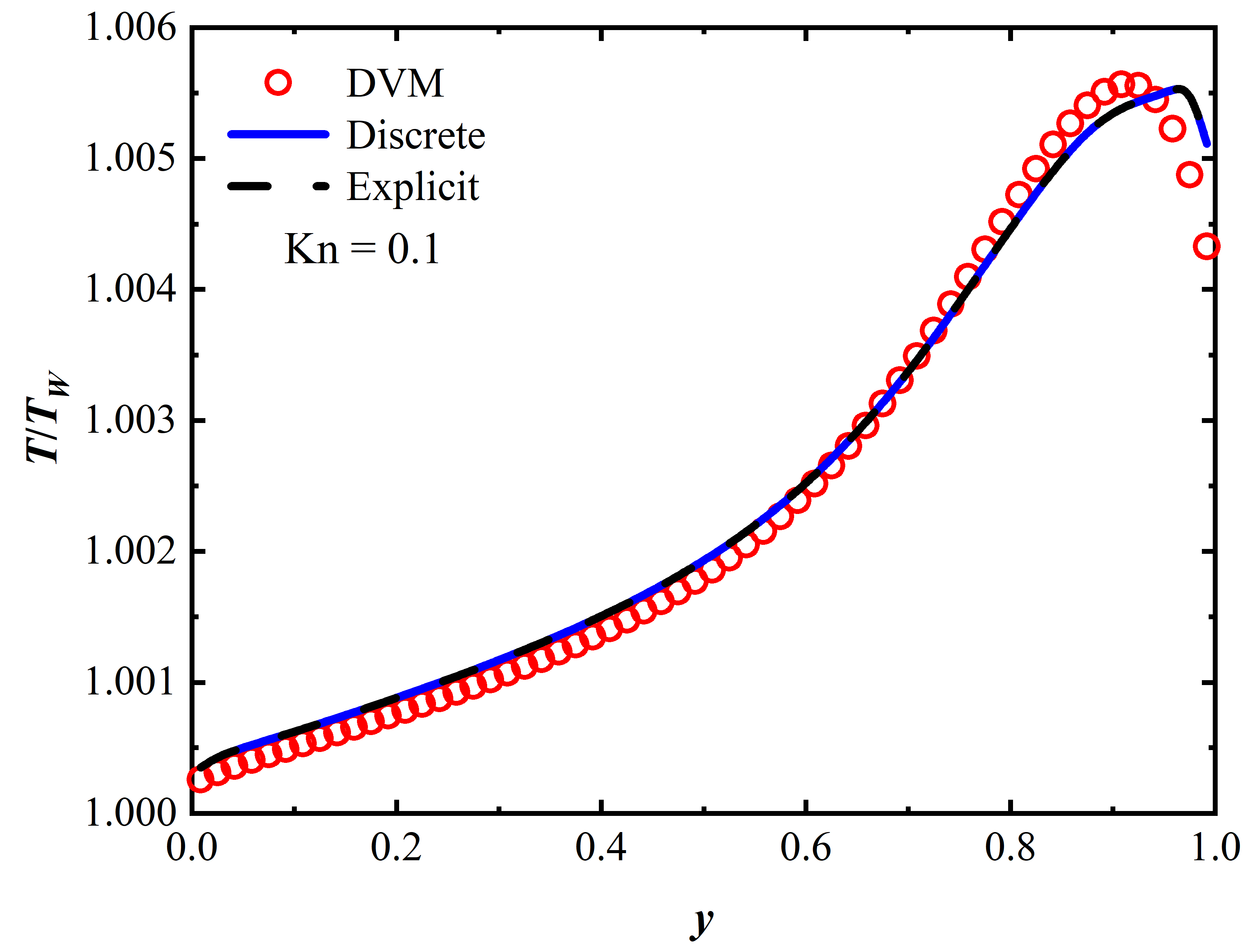}
\caption{Lid-driven cavity flow at $\mathrm{Kn}=0.1$. (Left) Velocity profiles along the central lines. (right) Temperature profile along the central line.}
\label{Fig10}
\end{figure}

\begin{figure}[H]
\centering
\includegraphics[width=7cm]{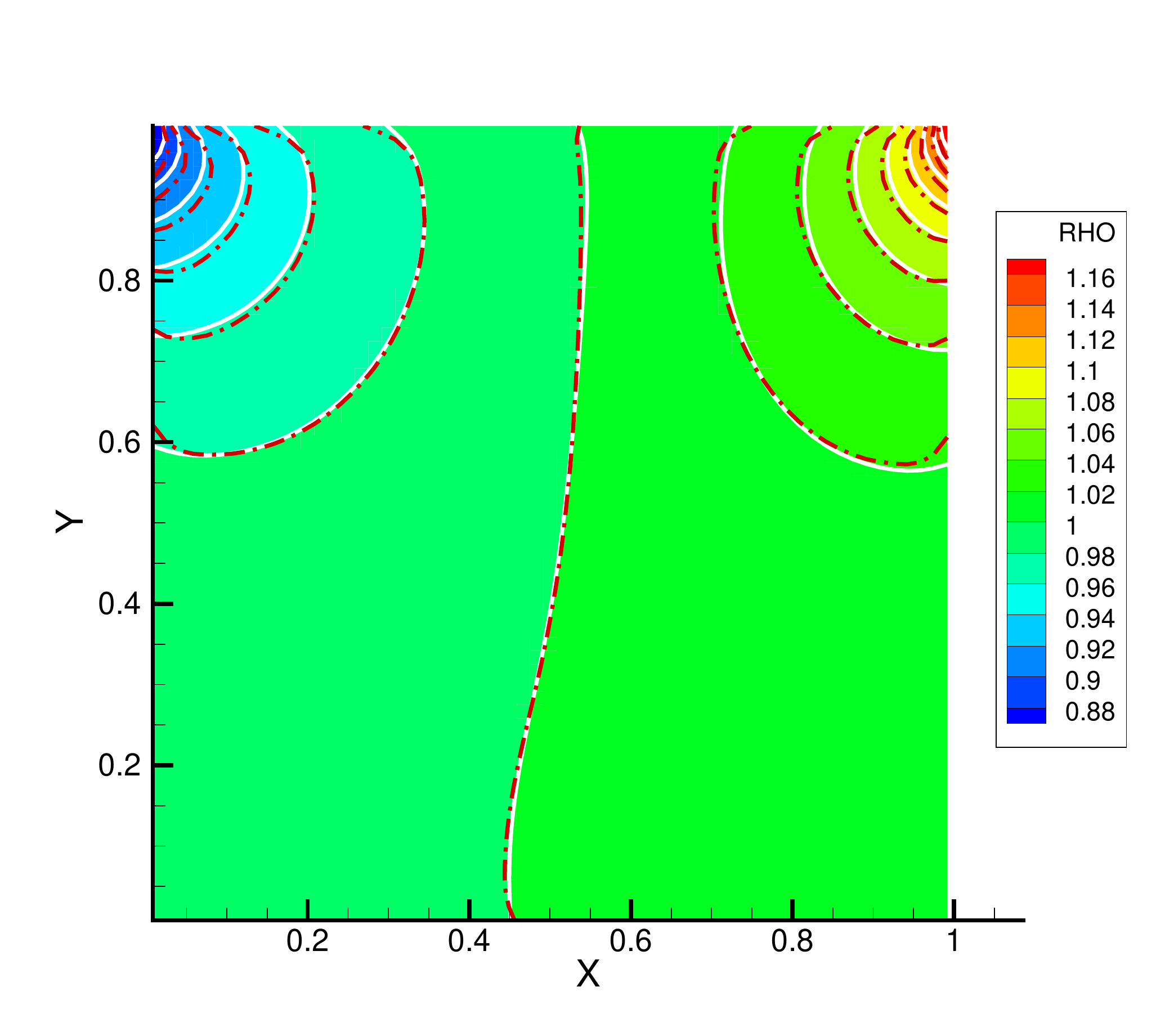}
\includegraphics[width=7cm]{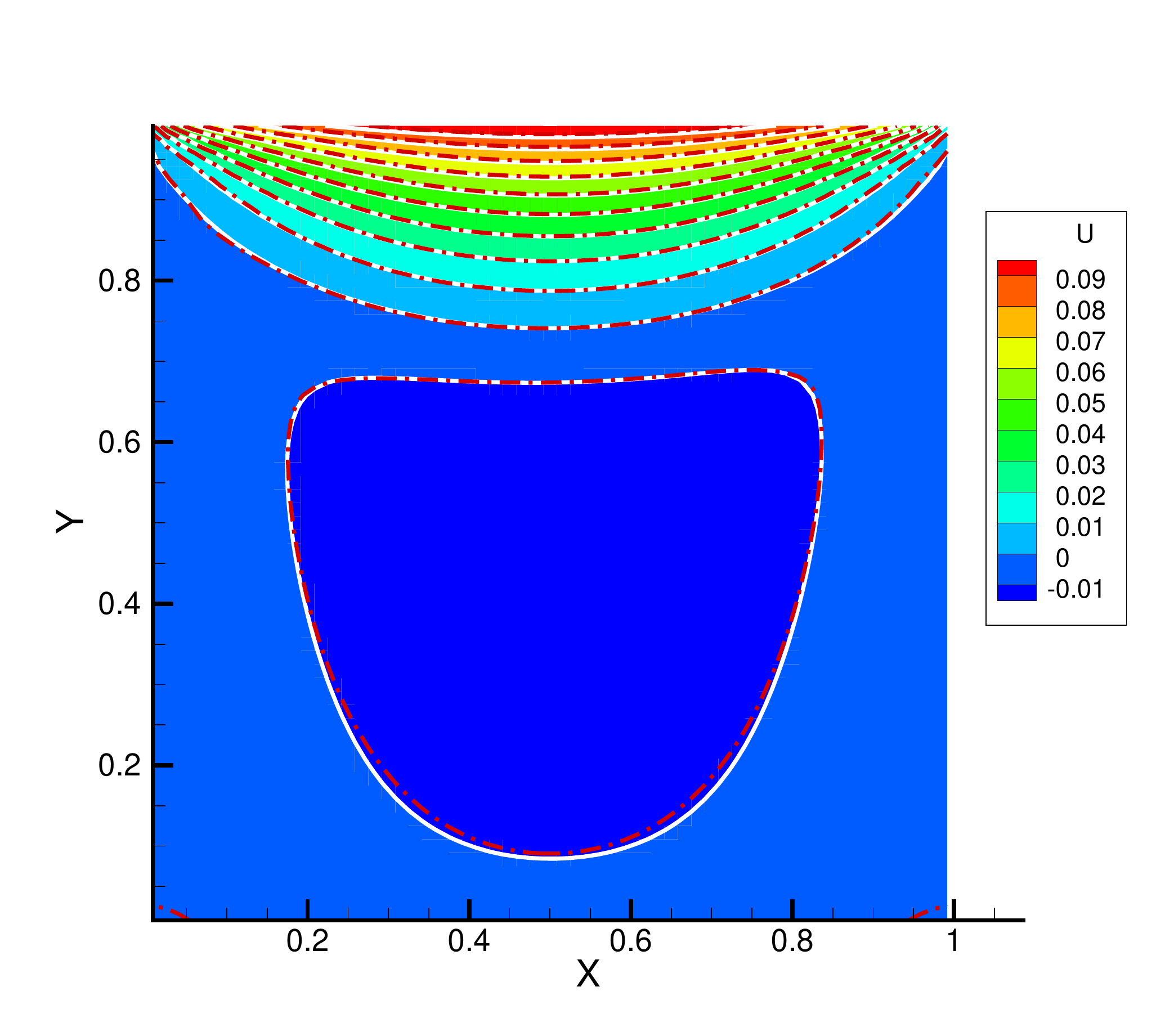}
\caption{Lid-driven cavity flow at $\mathrm{Kn}=0.1$. (Left) density  contours. (right) $U$-velocity contours. (Red dash dot line: explicit form of G13-MGKS; Colored background with white solid line: DVM)}
\label{Fig11}
\end{figure}

Further increase the Knudsen number to 0.1, the profiles of velocity and temperature are shown in Fig. \ref{Fig10}. Besides, the comparisons of density and $U$-velocity contours between the solutions from DVM and explicit form of G13-MGKS are displayed in Fig. \ref{Fig11}. It could be found that the results computed by the present method with the VDF of G13 could match well with the DVM in the whole flow field. Slight deviation appears in the temperature profile at the top area near the lid. The maximum relative error of the $T/T_{W}$ along the horizontal central line increases to 2.1\% at Kn = 0.1. This may be because the rarefied effect would become more significant near the wall.

\begin{figure}[H]
\centering
\includegraphics[width=7cm]{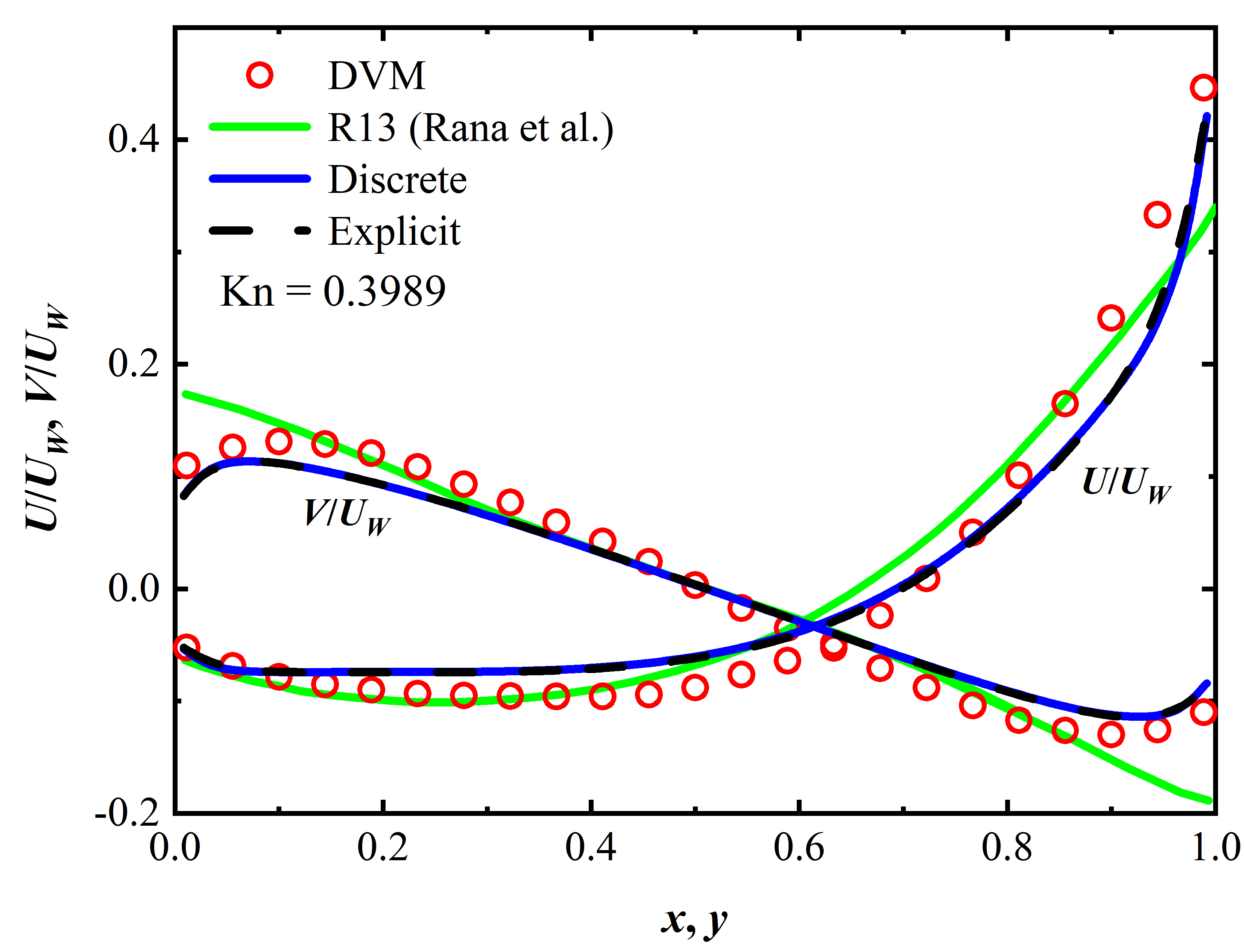}
\includegraphics[width=7cm]{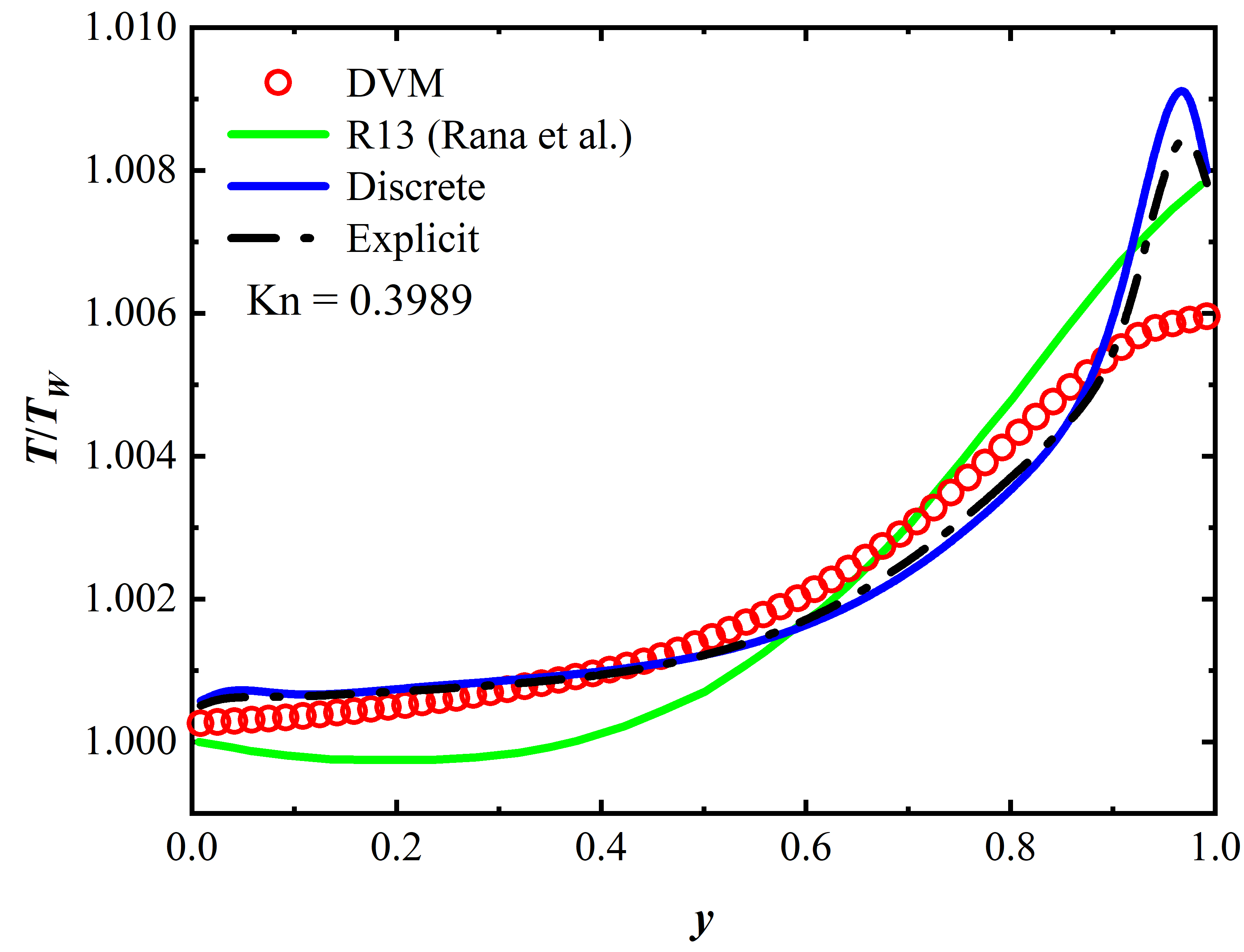}
\caption{Lid-driven cavity flow at $\mathrm{Kn}=0.3989$, (Left) Velocity profiles along the central lines and (right) Temperature profile along the central line.}
\label{Fig12}
\end{figure}

As the Knudsen number rises to Kn = 0.3989, the over-predicting of temperature at the top wall shown in Fig. \ref{Fig12} indicates that G13 loses its accuracy when a strong rarefied effect appears. Compared with the solutions from R13, the non-linear profiles of velocity can still be captured by the present method. It is interesting to note that the slight difference in temperature profile between the discrete and explicit form appears. In our test, increasing the number of discrete points in velocity space does not help to reduce this difference.   

\begin{figure}[H]
\centering
\includegraphics[width=7cm]{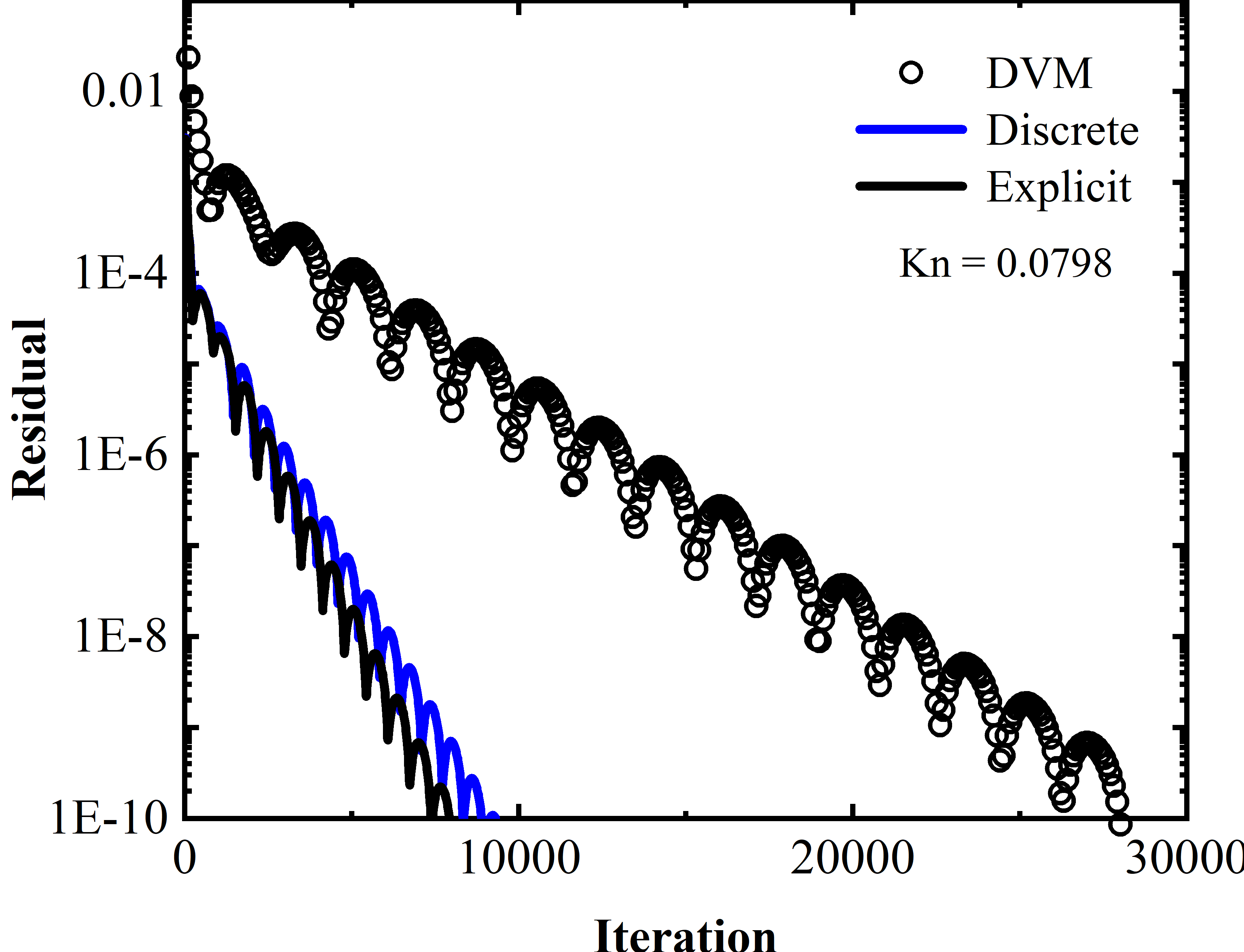}
\includegraphics[width=7cm]{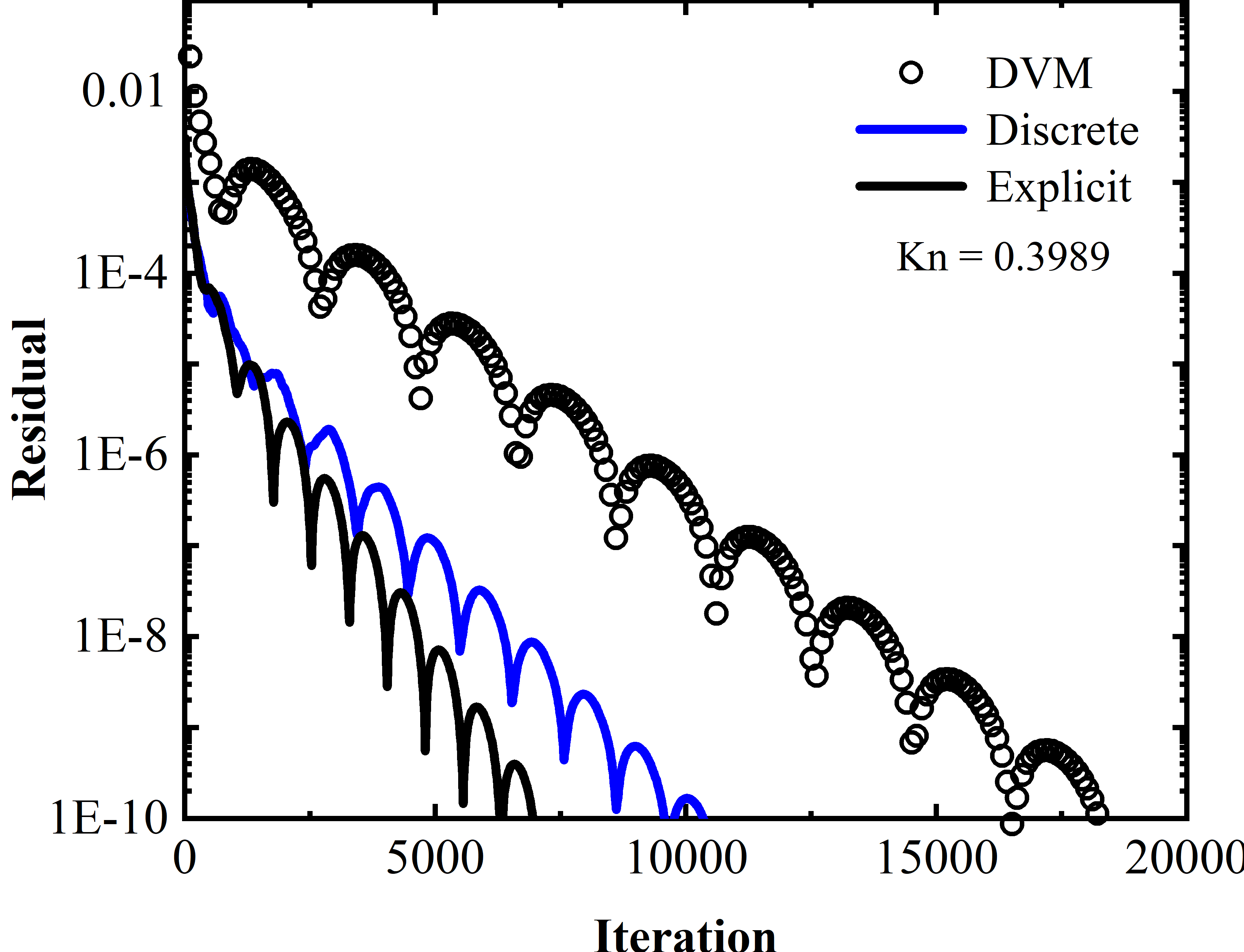}
\caption{Convergence history of Lid-driven cavity flows at (left) $\mathrm{Kn}=0.0798$ and (right) $\mathrm{Kn}=0.3989$.}
\label{Fig13}
\end{figure}

The comparisons of convergence history between discrete and explicit forms of G13-MGKS with DVM are given in Fig. \ref{Fig13}. Both the discrete and explicit forms converge quickly compared to DVM. Probably benefited from the avoidance of discretization in the molecular velocity space, better convergence appears in the explicit form. Moreover, the comparisons of computational times are shown in Table \ref{tb1} under different Knudsen numbers. A personal workstation with an Intel(R) Xeon(R) 4316 central processing unit (CPU) is tested with 9 threads of Open Multi-Processing (OpenMP) parallel computation. The results show that the discrete form of the present method consumes only about one-tenth of the computation time of DVM. This is because the G13 distribution function is a third-order polynomial with respect to peculiar velocities. Compared to the complex and unknown VDFs being evolved in the DVM, the truncated distribution function in lower-order polynomial form can be described using fewer velocity points (however, the ability to capture non-equilibrium effects is also restrained). Besides, the processing of distribution functions is restricted at the cell interface without manipulating a sizable number of discrete VDFs in the cell center. Moreover, the explicit form of the present method takes only a few tens of seconds and the computational times are only about one percent of the computational times of DVM. This demonstrates that eliminating the discretization in the molecular velocity space can dramatically improve computational efficiency.

\begin{table}[H]
\centering
\setlength{\abovecaptionskip}{10pt}
\setlength{\belowcaptionskip }{10pt}
\caption{Computational times (seconds) of different methods for the lid-driven cavity flow}
\setlength{\tabcolsep}{9mm}
\begin{threeparttable}
\begin{tabular}{cccc}
\hline
\hline
\textbf{Case}& \textbf{DVM}& \textbf{Discrete Form}& \textbf{Explicit Form}\\
\hline
Kn = 0.798    &   4221.6&   332.6&   22.2\\
Kn = 0.10    &   3959.1&   312.3&   24.2\\
Kn = 0.3989    &   2736.3&   437.5&   33.1\\
\hline
\hline
\end{tabular}
\label{tb1}
\end{threeparttable}
\end{table}

\subsection{\emph{Rayleigh Flow}}
\label{sec3-4}

\begin{figure}[H]
\centering
\includegraphics[width=7cm]{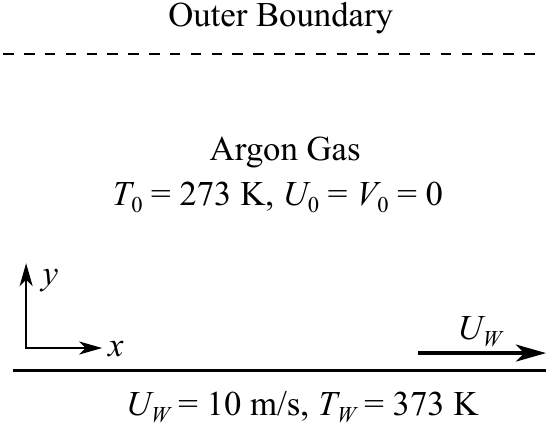}
\caption{Schematic of Rayleigh flow.}
\label{FigRayleigh}
\end{figure}

As shown in Fig. \ref{FigRayleigh}, The Rayleigh flow depicts an unsteady flow where a plate beneath a gas at rest suddenly obtains a constant parallel velocity of $U_{W}=10$ m/s with a constant temperature $T_{W}=373$ K. Following the setup from the work by Sun \cite{sun_information_2003}, the argon gas is at rest when $t=0$ with a temperature of $T_{0}=273$ K, molecular mass $R_{g}=208.13$ J/Kg K. After the plate moves when $t>0$, the shearing effect near the wall drives the gas field in an unsteady transport. The mean collision time of the Rayleigh flow is defined as $\tau_{0}=\lambda_{0} / \nu_{0}$ with the particle mean free path $\lambda_{0}$. The mean molecular speed denotes $v_{m}=2\sqrt{2 R_{g} T_{0} / \pi}$. The computational domain of $\left[-0.05, 0.05\right] \times\left[0, 1\right]$ is discretized by $10 \times 100$ cells uniformly. The top boundary is applied by the far-field boundary condition and the left and right sides of the computational domain are subjected to the periodic boundary condition. The Gauss–Hermite quadrature with $8 \times 8$ velocity points can be utilized in the velocity space of $\left[-4\sqrt{2 R T_{W}}, 4\sqrt{2 R T_{W}}\right]^{2}$ for the discrete form of G13-MGKS. For the purpose of comparison, the results from DVM with $28 \times 28$ Gauss–Hermite points are presented. 

\begin{figure}[H]
\centering
\includegraphics[width=7cm]{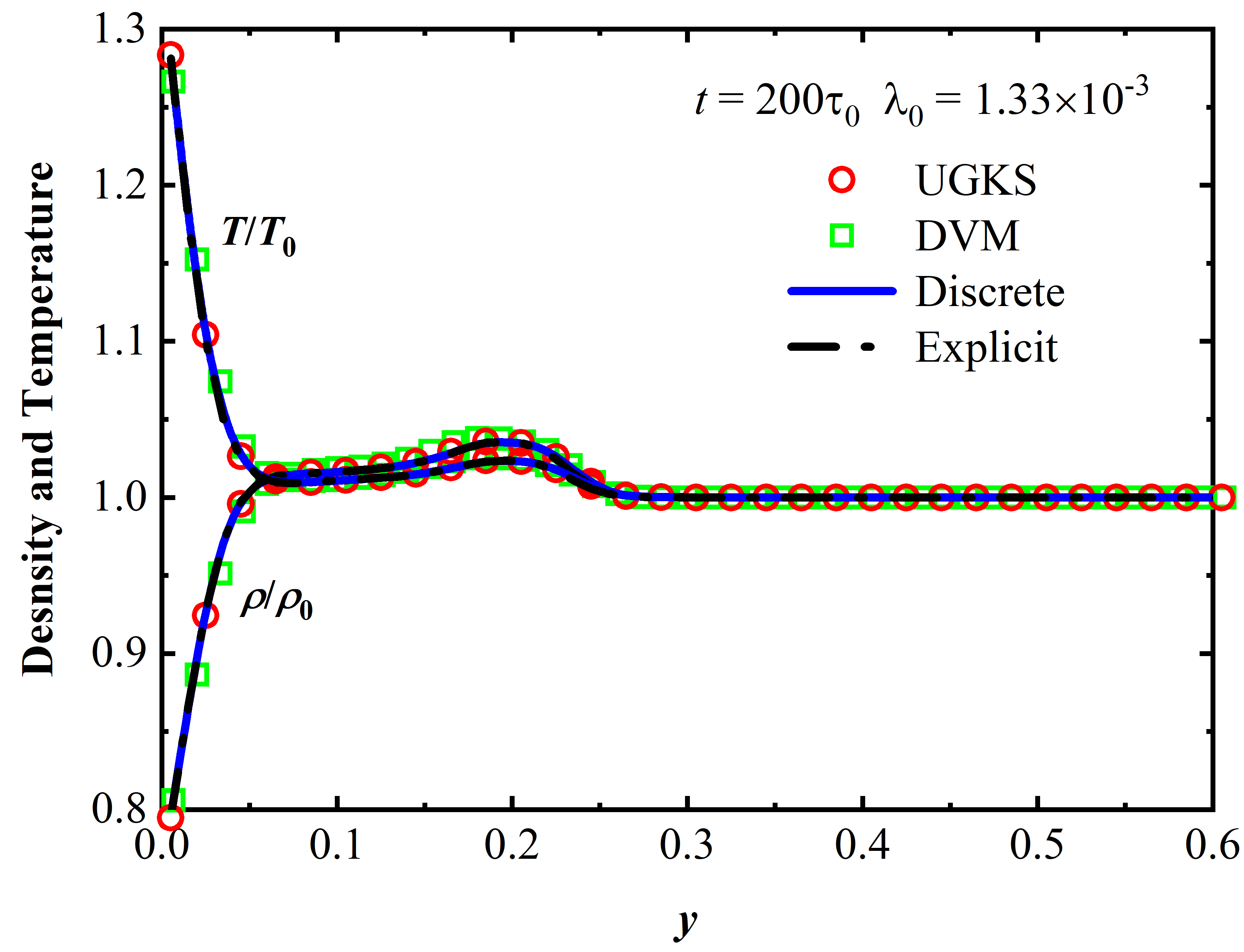}
\includegraphics[width=7cm]{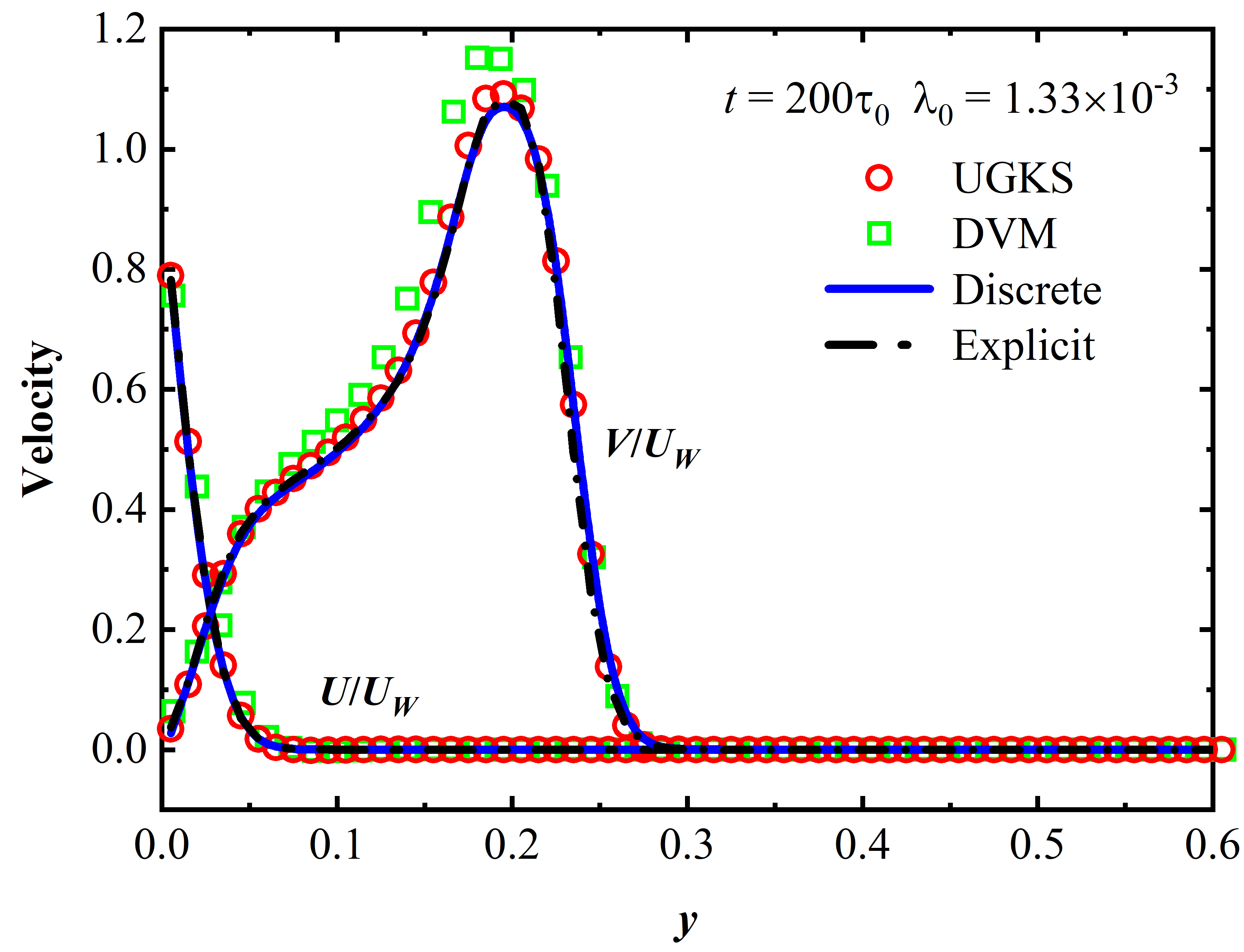}
\caption{Rayleigh flow at time $t=200\tau_{0}$, (left) Density and temperature, (right) $U$-velocity and $V$-velocity.}
\label{Fig14}
\end{figure}

\begin{figure}[H]
\centering
\includegraphics[width=7cm]{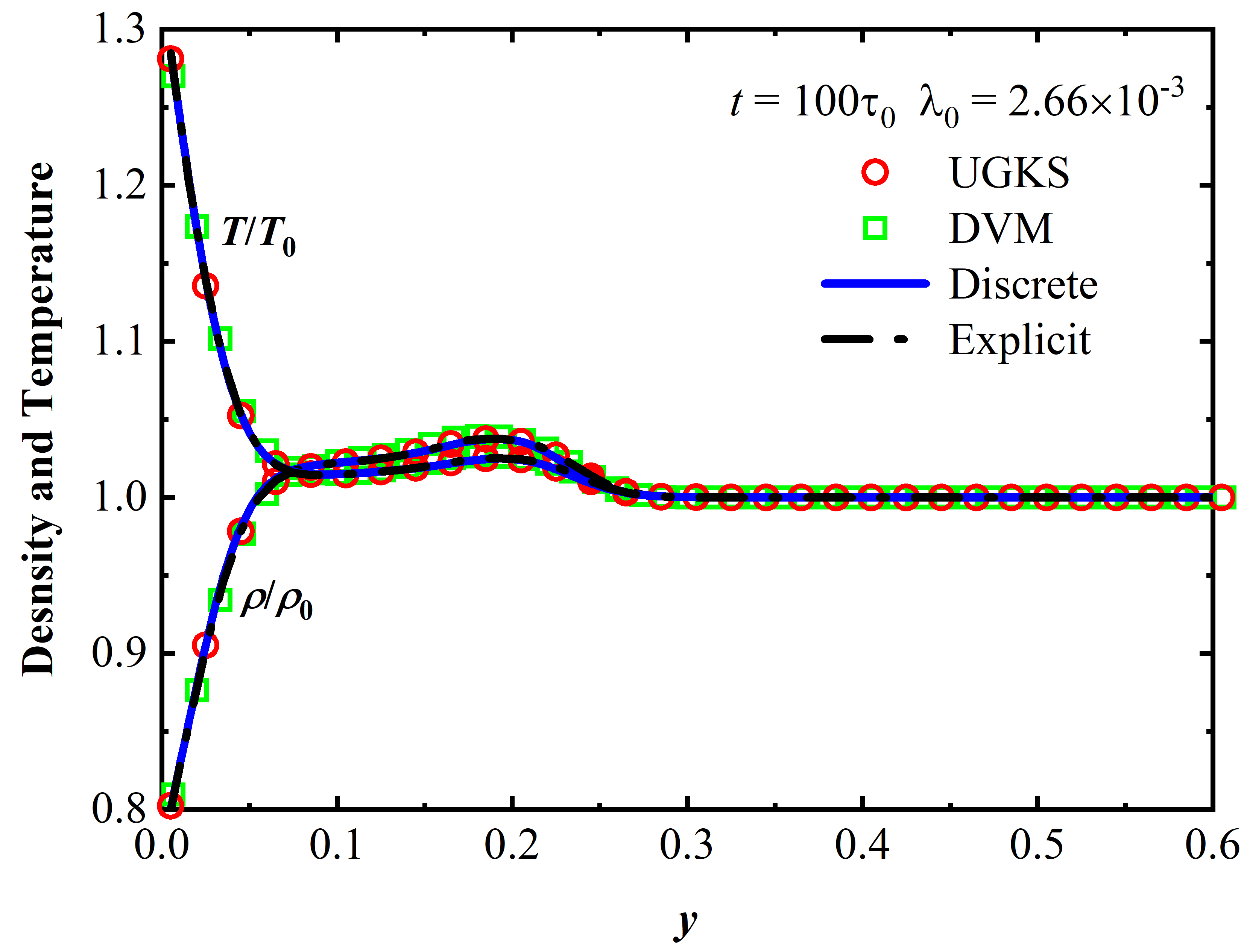}
\includegraphics[width=7cm]{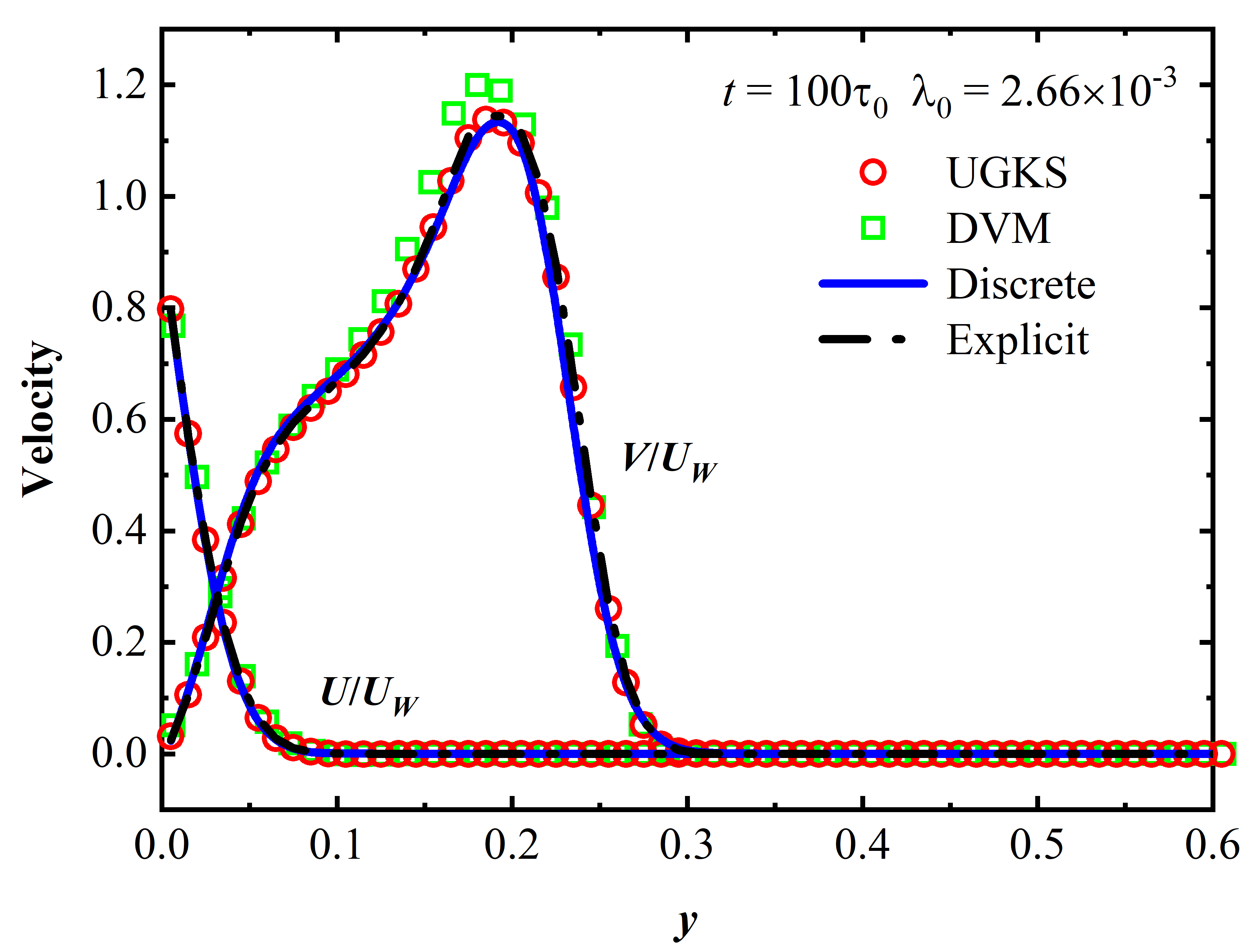}
\caption{Rayleigh flow at time $t=100\tau_{0}$, (left) Density and temperature, (right) $U$-velocity and $V$-velocity.}
\label{Fig15}
\end{figure}

Numerical simulations with a series of particle mean free paths are plotted in the present section. The solutions of normalized density, velocity and temperature at the time of $t=200\tau_{0}$ and $\lambda_{0}=1.33\times 10^{-3}$ are shown in Fig. \ref{Fig14}. The deviation between the solutions of UGKS and DVM appears in the normalized velocity $V/U_{W}$ apparently. This phenomenon was also found in a previous study of DVM \cite{huang_unified_2013} when the particle mean free path is dramatically smaller than cell size. At such condition, the DVM may depend on the cell size sensitively and cannot recover the hydrodynamic effect. Benefiting from collision effects being taken into account in the reconstruction of numerical fluxes as Eq. (\ref{eq12}), the solutions from the discrete and explicit forms of G13-MGKS agree well with the benchmark solutions from UGKS. When the particle mean free path increase to $\lambda_{0}=2.66\times 10^{-3}$ at time of $t=100\tau_{0}$, the discrete and explicit form of the G13-MGKS as shown in Fig. \ref{Fig15} still perform well, especially for the normalized velocity $V/U_{W}$ compared to DVM. 

\begin{figure}[H]
\centering
\includegraphics[width=7cm]{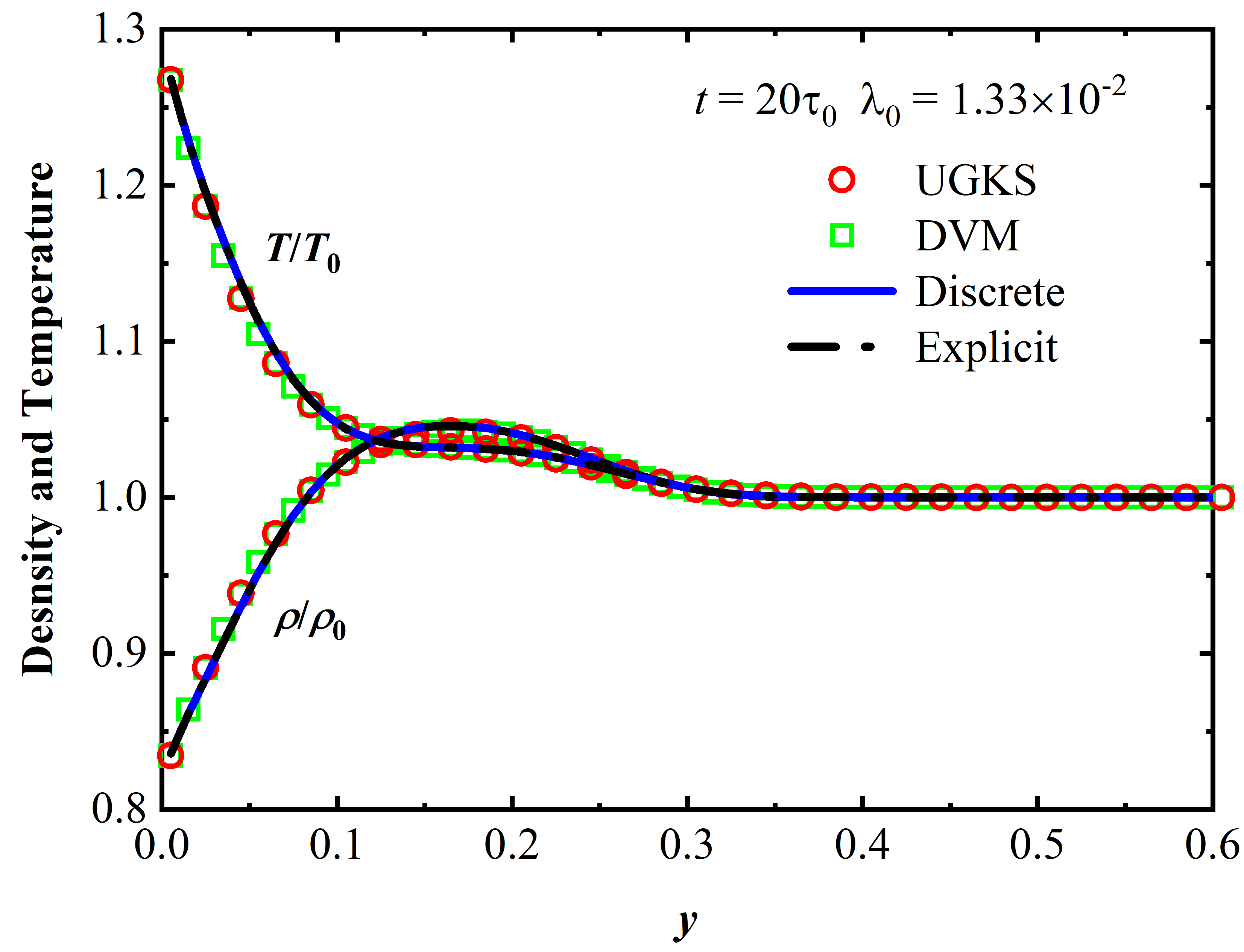}
\includegraphics[width=7cm]{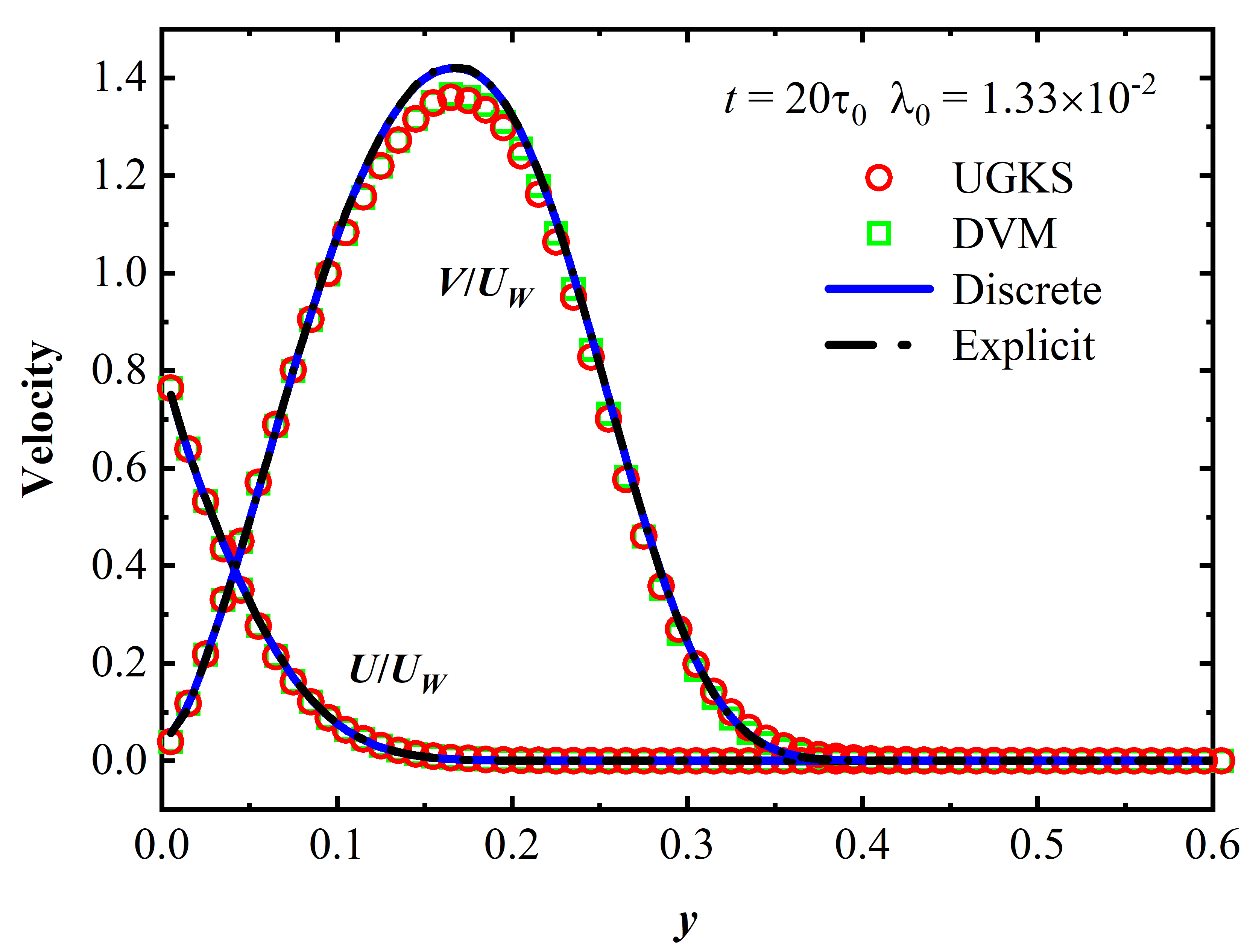}
\caption{Rayleigh flow at time $t=20\tau_{0}$, (left) Density and temperature, (right) $U$-velocity and $V$-velocity.}
\label{Fig16}
\end{figure}

\begin{figure}[H]
\centering
\includegraphics[width=7cm]{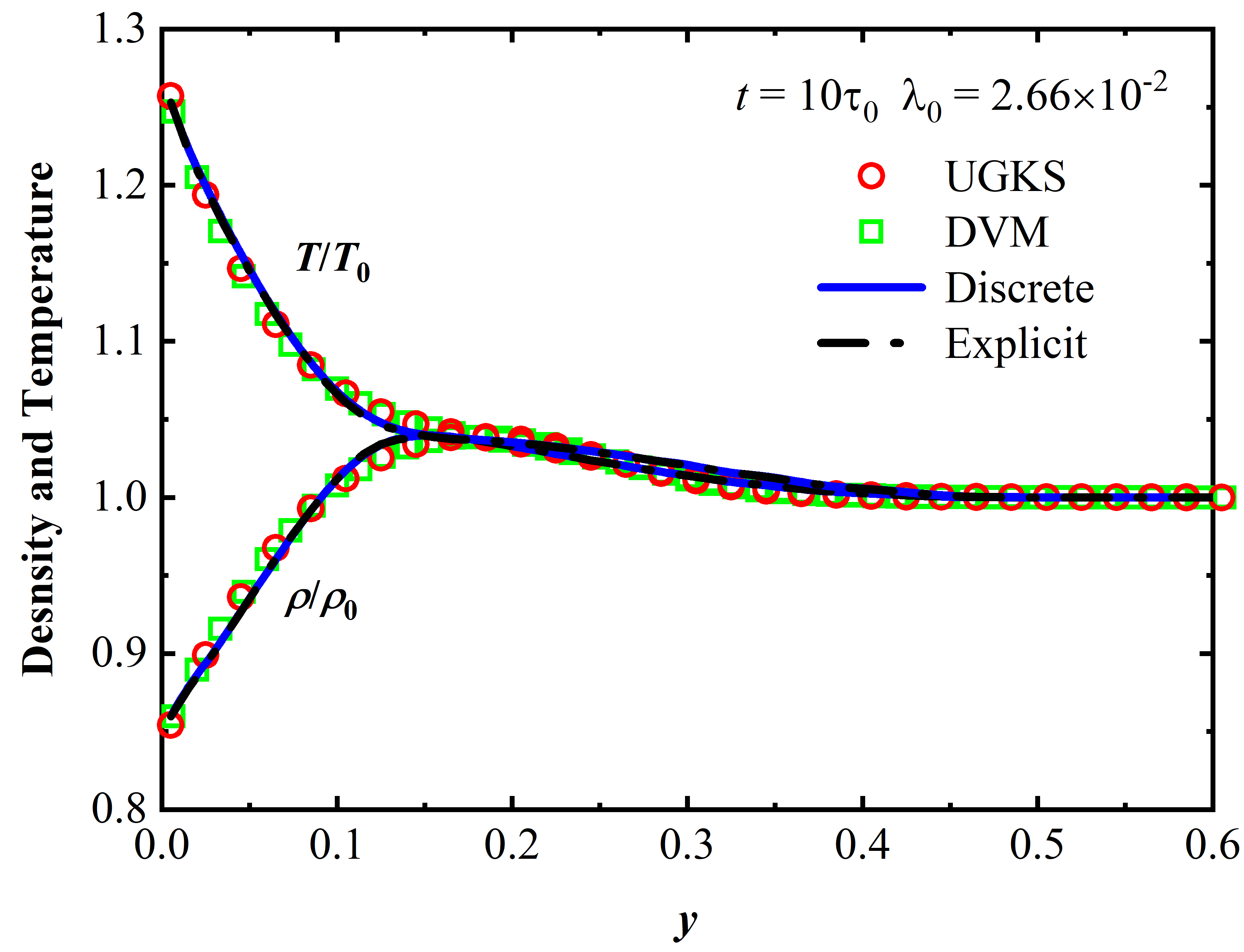}
\includegraphics[width=7cm]{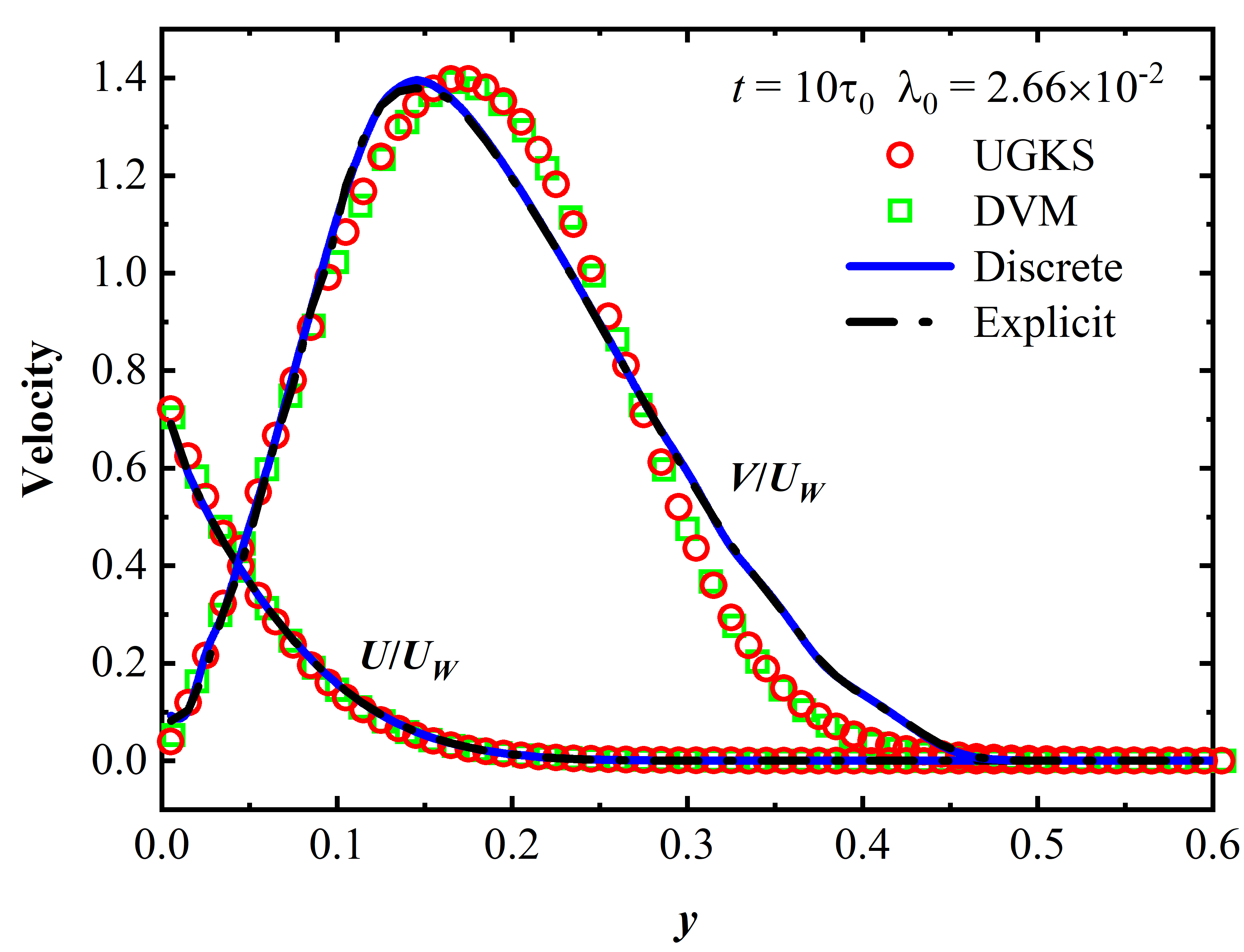}
\caption{Rayleigh flow at time $t=10\tau_{0}$, (left) Density and temperature, (right) $U$-velocity and $V$-velocity.}
\label{Fig17}
\end{figure}

As the particle mean free path rises to $\lambda_{0}=1.33\times 10^{-2}$, the solutions at the time of $t=20\tau_{0}$ are presented in Fig. \ref{Fig16}. It can be found that the deviation between the solutions of DVM and UGKS disappear. However, the present solver slightly overpredicts the maximum value of the normalized velocity $V/U_{W}$ compared to UGKS and DVM. The relative error about the maximum normalized velocity is about 3.96\%. Further increasing the particle mean free path to $\lambda_{0}=2.66\times 10^{-2}$, we can find the obvious deviation between the solutions of present solver and the reference results in Fig. \ref{Fig17}. This phenomenon may be induced by the non-equilibrium effect in a larger region of the Knudsen layer. Higher-order distribution functions or hybrid methods may help to alleviate this issue. 

Table \ref{tb2} presents the computational costs of the different methods. Based on the present data, it can be found that the discrete and explicit form of the present solver is 3.08 and 22.7 times faster than the DVM. The memory consumption of the discrete and explicit from of the present solver cost about 27.5\% and 8.5\% of the DVM, respectively. From the current numerical experiments, the present framework performs more efficiently and is less memory demanding for the simulations. 

\begin{table}[H]
\centering
\setlength{\abovecaptionskip}{10pt}
\setlength{\belowcaptionskip }{10pt}
\caption{Computational times and memory consumption of different methods for Rayleigh flow.}
\setlength{\tabcolsep}{9mm}
\begin{threeparttable}
\begin{tabular}{cccc}
\hline
\hline
\textbf{}& \textbf{DVM}& \textbf{Discrete Form}& \textbf{Explicit Form}\\
\hline
\textbf{Time}& 65.8 s&   21.3 s&   2.9 s\\
\textbf{Memory}& 71.3 MB&   19.6 MB&   6 MB\\
\hline
\hline
\end{tabular}
\label{tb2}
\end{threeparttable}
\end{table}

\section{CONCLUSIONS}
\label{S:4}

To efficiently simulate multiscale flows by the macroscopic equations, the G13 based Moment Gas Kinetic scheme (G13-MGKS) is proposed in the present work. The macroscopic equations of the stress and heat flux are derived from the moment integral of the discrete Boltzmann equation in the finite volume framework. The VDF at cell interface can be constructed by the Boltzmann-BGK equation along the characteristic line and the numerical fluxes are given as the discrete form and explicit form. The intricate partial differential for the high order moments and discretization of molecular velocity space can be avoided, so the present framework incorporates simplicity and efficiency. 

To evaluate the performance of the G13 with the present framework, Four numerical examples including the shock wave structure, Sod shock tube problem, the lid-driven cavity flow and the unsteady Rayleigh flow are tested covering the steady and unsteady, microscale and supersonic rarefied flow. Based on the comparisons of G13-MGKS with different methods, the G13-MGKS performs the accurate solutions in the continuum flow regime and can predict reasonable solutions for flows in the rarefied regime. The numerical test about the convergence history and the computational efficiency indicates the good stability of G13-MGKS. The present framework with the discrete form would be helpful for the researcher to apply and evaluate novel distribution functions of numerical fluxes. The superior efficiency of the present method with the explicit expression of numerical fluxes retain the potential for practical engineering applications. 

\section*{ACKNOWLEDGMENTS}
\label{S:5}

This work was partly supported by the Ministry of Education (MOE) of Singapore (Grant No. MOE2018-T2-1-135). We acknowledge the computing support provided by the National Super-computing Centre (NSCC) Singapore for Computational Science and the High-Performance Computing of NUS (NUS-HPC).

\appendix
\section{PARAMETERS FOR COMPUTATION OF $\left\langle\xi_{n}^{o} \xi_{\tau}^{p} \zeta^{q} f_{i j}\right\rangle_{>0}$}
\label{A:1}

Here we take the integration conducted at the left side of cell interface as an example and  we only need to replace the $\langle\cdot\rangle_{>0}$ by the $\langle\cdot\rangle_{<0}$ for the integration at right side. The parameters of $\left\langle\xi_{n}^{o} \xi_{\tau}^{p} \zeta^{q} f_{i j}\right\rangle_{>0}$ could be calculated using binomial theory as follows:

\begin{equation}
\begin{aligned}
\left\langle\xi_{n}^{o} \xi_{\tau}^{p} \zeta^{q} f_{i j}\right\rangle_{>0}=\sum_{m=0}^{p} \frac{p !}{m !(p-m) !}\left\langle\xi_{n}^{o} C_{\tau}^{m} \zeta^{q} f_{i j}\right\rangle_{>0}\left(U_{n}\right)^{p-m},
\end{aligned}
\label{eqA1}
\end{equation}

To facilitate the description, the notation of the integration of Maxwellian equilibrium state is written as $\langle\cdot\rangle^{e q}=\int_{-\infty}^{+\infty} \cdot g d \bm{\xi}$. Then, the moment of VDF $\left\langle\xi_{n}^{o} C_{\tau}^{p} \zeta^{q} f_{i j}\right\rangle_{>0}$ could computed as

\begin{equation}
\begin{aligned}
\left\langle\xi_{n}^{o} C_{\tau}^{p} \zeta^{q} f_{i j}\right\rangle_{>0}=&\alpha \bigg( \beta\left\langle\xi_{n}^{o} C_{n}^{0}\right\rangle_{>0}^{eq}\left\langle C_{\tau}^{p}\right\rangle^{eq}\left\langle\zeta^{q}\right\rangle^{eq}+\sigma_{n n}^{*}\left\langle\xi_{n}^{o} C_{n}^{2}\right\rangle_{>0}^{eq}\left\langle C_{\tau}^{p}\right\rangle^{eq}\left\langle\zeta^{q}\right\rangle^{eq} \\
&+2 \sigma_{n \tau}^{*}\left\langle\xi_{n}^{o} C_{n}^{1}\right\rangle_{>0}^{eq}\left\langle C_{\tau}^{p+1}\right\rangle^{eq}\left\langle\zeta^{q}\right\rangle^{eq}+\sigma_{\tau \tau}^{*}\left\langle\xi_{n}^{o} C_{n}^{0}\right\rangle_{>0}^{eq}\left\langle C_{\tau}^{p+2}\right\rangle^{eq}\left\langle\zeta^{q}\right\rangle^{eq} \\
&-\left(\sigma_{n n}^{*}+\sigma_{\tau \tau}^{*}\right)\left\langle\xi_{n}^{o} C_{n}^{0}\right\rangle_{>0}^{eq}\left\langle C_{\tau}^{p}\right\rangle^{eq}\left\langle\zeta^{q+2}\right\rangle^{eq} \\
&-q_{n}^{*}\left\langle\xi_{n}^{o} C_{n}^{1}\right\rangle_{>0}^{eq}\left\langle C_{\tau}^{p}\right\rangle^{eq}\left\langle\zeta^{q}\right\rangle^{eq}-q_{\tau}^{*}\left\langle\xi_{n}^{o} C_{n}^{0}\right\rangle_{>0}^{eq}\left\langle C_{\tau}^{p+1}\right\rangle^{eq}\left\langle\zeta^{q}\right\rangle^{eq} \\
&+0.4 \lambda q_{n}^{*}\left(\left\langle\xi_{n}^{o} C_{n}^{3}\right\rangle_{>0}^{eq}\left\langle C_{\tau}^{p}\right\rangle^{eq}\left\langle\zeta^{q}\right\rangle^{eq}+\left\langle\xi_{n}^{o} C_{n}^{1}\right\rangle_{>0}^{eq}\left\langle C_{\tau}^{p+2}\right\rangle^{eq}\left\langle\zeta^{q}\right\rangle^{eq}\right.\\
&\left.+\left\langle\xi_{n}^{o} C_{n}^{1}\right\rangle_{>0}^{eq}\left\langle C_{\tau}^{p}\right\rangle^{eq}\left\langle\zeta^{q+2}\right\rangle^{eq}\right) \\
&+0.4 \lambda q_{\tau}^{*}\left(\left\langle\xi_{n}^{o} C_{n}^{2}\right\rangle_{>0}^{eq}\left\langle C_{\tau}^{p+1}\right\rangle^{eq}\left\langle\zeta^{q}\right\rangle^{eq}\right.\\
&\left.+\left\langle\xi_{n}^{o} C_{n}^{0}\right\rangle_{>0}^{eq}\left\langle C_{\tau}^{p+3}\right\rangle^{eq}\left\langle\zeta^{q}\right\rangle^{eq}+\left\langle\xi_{n}^{o} C_{n}^{0}\right\rangle_{>0}^{eq}\left\langle C_{\tau}^{p+1}\right\rangle^{eq}\left\langle\zeta^{q+2}\right\rangle^{eq}\right) \bigg),
\end{aligned}
\label{eqA2}
\end{equation}

\noindent where $\alpha=1-\frac{\Delta t}{\tau}$ and $\beta=\frac{\tau}{\tau-\Delta t}$ are the coefficients related the time terms. The terms of stress and heat flux marked with an asterisk superscript are given by

\begin{equation}
\sigma_{x x}^{*}=\sigma_{x x} /(2 p R T), \quad \sigma_{x y}^{*}=\sigma_{x y} /(2 p R T), \quad \sigma_{y y}^{*}=\sigma_{y y} /(2 p R T),
\label{eqA3}
\end{equation}

\noindent and

\begin{equation}
q_{x}^{*}=q_{x} /(p R T), \quad q_{y}^{*}=q_{y} /(p R T).
\label{eqA4}
\end{equation}

The next task is to calculate term $\left\langle\xi_{n}^{p} C_{n}^{q}\right\rangle_{>0}^{eq}$, $\left\langle C_{\tau}^{k}\right\rangle^{eq}$ and $\left\langle\zeta^{k}\right\rangle^{eq}$, which are related to the moment integration of equilibrium state. Also from the binomial theory, the $\left\langle\xi_{n}^{p} C_{n}^{q}\right\rangle_{>0}^{eq}$ could be expressed by the linear combination of $\left\langle\xi_{n}^{k}\right\rangle_{>0}^{eq}$ as

\begin{equation}
\left\langle\xi_{n}^{p} C_{n}^{q}\right\rangle_{>0}^{eq}=(-1)^{q-m} \sum_{m=0}^{n} \frac{q !}{m !(q-m) !}\left\langle\xi_{n}^{m+p}\right\rangle_{>0}^{e q}\left(U_{n}^{L}\right)^{q-m}.
\label{eqA5}
\end{equation}

Considering that the expression of $\left\langle\xi_{n}^{k}\right\rangle_{>0}^{eq}$ and $\left\langle\xi_{n}^{k}\right\rangle_{<0}^{eq}$ have different manner, the integration parameters related to the equilibrium state $\left\langle C_{\tau}^{k}\right\rangle^{eq}$, $\left\langle\zeta^{k}\right\rangle^{eq}$, $\left\langle\xi_{n}^{k}\right\rangle_{>0}^{eq}$ and $\left\langle\xi_{n}^{k}\right\rangle_{<0}^{eq}$ given in \ref{B:1}.

\section{COMPUTATION OF INTEGRATION PARAMETERS RELATED TO THE EQUILIBRIUM STATE}
\label{B:1}

Taking the notation of integral from zero to infinite on the left side of cell interface, the integration parameter $\left\langle\xi_{n}^{k}\right\rangle_{>0}^{eq}$ could be given as

\begin{equation}
\left\langle\xi_{n}^{0}\right\rangle_{>0}^{e q}=\frac{1}{2}\left[1+\operatorname{erf}\left(\sqrt{\lambda^{L}} U_{n}^{L}\right)\right],
\label{eqB1}
\end{equation}

\begin{equation}
\left\langle\xi_{n}^{1}\right\rangle_{>0}^{e q}=U_{n}^{L}\left\langle\xi_{n}^{0}\right\rangle_{>0}^{e q}+\frac{1}{2} \frac{e^{-\lambda^{L}\left(U_{n}^{L}\right)^{2}}}{\sqrt{\lambda^{L} \pi}},
\label{eqB2}
\end{equation}

\begin{equation}
\left\langle\xi_{n}^{k+2}\right\rangle_{>0}^{e q}=U_{n}^{L}\left\langle\xi_{n}^{k+1}\right\rangle_{>0}^{e}+\frac{k+1}{2 \lambda^{L}}\left\langle\xi_{n}^{k}\right\rangle_{>0}^{e q}, \quad k=0,1,2, \ldots,
\label{eqB3}
\end{equation}

\noindent where $\lambda=1 /\left(2 R_{g} T\right)$. Similarly, taking the notation of integral from negative infinite to zero, the integration parameter $\left\langle\xi_{n}^{k}\right\rangle_{<0}^{eq}$ are 

\begin{equation}
\left\langle\xi_{n}^{0}\right\rangle_{<0}^{e q}=\frac{1}{2} \operatorname{erfc}\left(\sqrt{\lambda^{R}} U_{n}^{R}\right),
\label{eqB4}
\end{equation}

\begin{equation}
\left\langle\xi_{n}^{1}\right\rangle_{<0}^{e q}=U_{n}^{R}\left\langle\xi_{n}^{0}\right\rangle_{<0}^{e q}-\frac{1}{2} \frac{e^{-\lambda^{R}\left(U_{n}^{R}\right)^{2}}}{\sqrt{U_{n}^{R} \pi}},
\label{eqB5}
\end{equation}

\begin{equation}
\left\langle\xi_{n}^{k+2}\right\rangle_{<0}^{e q}=U_{n}^{R}\left\langle\xi_{n}^{k+1}\right\rangle_{<0}^{e q}+\frac{k+1}{2 \lambda^{R}}\left\langle\xi_{n}^{k}\right\rangle_{<0}^{e q}, \quad k=0,1,2, \ldots,
\label{eqB6}
\end{equation}

Following the binomial theory, part of even order of integration parameters $\left\langle C_{\tau}^{k}\right\rangle^{eq}$ and $\left\langle\zeta^{k}\right\rangle^{eq}$ could be computed as

\begin{equation}
\left\langle C_{\tau}^{0}\right\rangle^{e q}=\left\langle\zeta^{0}\right\rangle^{e q}=1,
\label{eqB7}
\end{equation}

\begin{equation}
\left\langle C_{\tau}^{2}\right\rangle^{e q}=\left\langle\zeta^{2}\right\rangle^{e q}=\frac{1}{2 \lambda},
\label{eqB8}
\end{equation}

\begin{equation}
\left\langle C_{\tau}^{4}\right\rangle^{e q}=\left\langle\zeta^{4}\right\rangle^{e q}=\frac{3}{4 \lambda^{2}},
\label{eqB9}
\end{equation}

\begin{equation}
\left\langle C_{\tau}^{6}\right\rangle^{e q}=\left\langle\zeta^{6}\right\rangle^{e q}=\frac{15}{8 \lambda^{3}},
\label{eqB10}
\end{equation}

\begin{equation}
\left\langle C_{\tau}^{8}\right\rangle^{e q}=\left\langle\zeta^{8}\right\rangle^{e q}=\frac{105}{16 \lambda^{4}}.
\label{eqB11}
\end{equation}

\noindent When $n$ is odd, the moment integrals of $\left\langle C_{\tau}^{k}\right\rangle^{eq}$ and $\left\langle\zeta^{k}\right\rangle^{eq}$ are all zero.

\section{COMPUTATION OF NUMERICAL FLUXES RELATED TO THE STRESS AND HEAT FLUX}
\label{C:1}

The formulations of parameters including $\mathbf{B}(1) \sim \mathbf{B}(4)$, $\mathbf{B}_{n}(1) \sim \mathbf{B}_{n}(4)$ and $\mathbf{B}_{\tau}(1) \sim \mathbf{B}_{\tau}(4)$ can be computed as

\begin{equation}
\mathbf{B}^{L}(1)=\mathbf{A}_{n}^{L}(2), \mathbf{B}_{n}^{L}(1)=\left\langle\xi_{n}^{4} \xi_{\tau}^{0} \zeta^{0} f_{i j}\right\rangle_{>0}, \mathbf{B}_{\tau}^{L}(1)=\left\langle\xi_{n}^{3} \xi_{\tau}^{1} \zeta^{0} f_{i j}\right\rangle_{>0},
\label{eqC1}
\end{equation}

\begin{equation}
\mathbf{B}^{L}(2)=\mathbf{A}_{\tau}^{L}(2), \mathbf{B}_{n}^{L}(2)=\mathbf{B}_{\tau}^{L}(1), \mathbf{B}_{\tau}^{L}(2)=\left\langle\xi_{n}^{2} \xi_{\tau}^{2} \zeta^{0} f_{i j}\right\rangle_{>0},
\label{eqC2}
\end{equation}

\begin{equation}
\mathbf{B}^{L}(3)=\mathbf{A}_{\tau}^{L}(3), \mathbf{B}_{n}^{L}(3)=\mathbf{B}_{\tau}^{L}(2), \mathbf{B}_{\tau}^{L}(3)=\left\langle\xi_{n}^{1} \xi_{\tau}^{3} \zeta^{0} f_{i j}\right\rangle_{>0},
\label{eqC3}
\end{equation}

\begin{equation}
\mathbf{B}^{L}(4)=\left\langle\xi_{n}^{1} \xi_{\tau}^{0} \zeta^{2} f_{i j}\right\rangle_{>0}, \mathbf{B}_{n}^{L}(4)=\left\langle\xi_{n}^{2} \xi_{\tau}^{0} \zeta^{2} f_{i j}\right\rangle_{>0}, \mathbf{B}_{\tau}^{L}(4)=\left\langle\xi_{n}^{1} \xi_{\tau}^{1} \zeta^{2} f_{i j}\right\rangle_{>0},
\label{eqC4}
\end{equation}

The formulations of parameters including $\mathbf{C}(1) \sim \mathbf{C}(6)$, $\mathbf{C}_{n}(1) \sim \mathbf{C}_{n}(6)$ and $\mathbf{C}_{\tau}(1) \sim \mathbf{C}_{\tau}(6)$ can be computed as

\begin{equation}
\mathbf{C}^{L}(1)=\mathbf{B}_{n}^{L}(1), \mathbf{C}_{n}^{L}(1)=\left\langle\xi_{n}^{5} \xi_{\tau}^{0} \zeta^{0} f_{i j}\right\rangle_{>0}, \mathbf{C}_{\tau}^{L}(1)=\left\langle\xi_{n}^{4} \xi_{\tau}^{1} \zeta^{0} f_{i j}\right\rangle_{>0},
\label{eqC5}
\end{equation}

\begin{equation}
\mathbf{C}^{L}(2)=\mathbf{B}_{n}^{L}(3), \mathbf{C}_{n}^{L}(2)=\left\langle\xi_{n}^{3} \xi_{\tau}^{2} \zeta^{0} f_{i j}\right\rangle_{>0}, \mathbf{C}_{\tau}^{L}(2)=\left\langle\xi_{n}^{2} \xi_{\tau}^{3} \zeta^{0} f_{i j}\right\rangle_{>0},
\label{eqC6}
\end{equation}

\begin{equation}
\mathbf{C}^{L}(3)=\mathbf{B}_{n}^{L}(4), \mathbf{C}_{n}^{L}(3)=\left\langle\xi_{n}^{3} \xi_{\tau}^{0} \zeta^{2} f_{i j}\right\rangle_{>0}, \mathbf{C}_{\tau}^{L}(3)=\left\langle\xi_{n}^{2} \xi_{\tau}^{1} \zeta^{2} f_{i j}\right\rangle_{>0},
\label{eqC7}
\end{equation}

\begin{equation}
\mathbf{C}^{L}(4)=\mathbf{B}_{n}^{L}(2), \mathbf{C}_{n}^{L}(4)=\mathbf{C}_{\tau}^{L}(1), \mathbf{C}_{\tau}^{L}(4)=\mathbf{C}_{n}^{L}(2),
\label{eqC8}
\end{equation}

\begin{equation}
\mathbf{C}^{L}(5)=\mathbf{B}_{\tau}^{L}(3), \mathbf{C}_{n}^{L}(5)=\mathbf{C}_{\tau}^{L}(2), \mathbf{C}_{\tau}^{L}(5)=\left\langle\xi_{n}^{1} \xi_{\tau}^{4} \zeta^{0} f_{i j}\right\rangle_{>0},
\label{eqC9}
\end{equation}

\begin{equation}
\mathbf{C}^{L}(6)=\mathbf{B}_{\tau}^{L}(4), \mathbf{C}_{n}^{L}(6)=\mathbf{C}_{\tau}^{L}(3), \mathbf{C}_{\tau}^{L}(6)=\left\langle\xi_{n}^{1} \xi_{\tau}^{2} \zeta^{2} f_{i j}\right\rangle_{>0},
\label{eqC10}
\end{equation}


\bibliographystyle{SG13-GKFS.bst}
\bibliography{SG13-GKFS.bib}
\end{document}